\documentclass[review]{elsarticle}

\usepackage{lineno,hyperref}
\usepackage{amsmath}
\usepackage{csquotes}
\MakeOuterQuote{"}
\usepackage{bm}
\usepackage{pgfplots}
\usepackage{tikz}
\usepackage[capitalise]{cleveref}  % Clever references to stuff.
\usepackage[acronym]{glossaries-extra}

\modulolinenumbers[5]

\journal{Computer Physics Communications}
\crefname{figure}{Fig.}{Figs.}
\crefname{equation}{Eq.}{Eqs.}
\crefname{section}{Sec.}{Secs.}

\glsdisablehyper
\setabbreviationstyle[acronym]{long-short}
\newacronym{ep}{EP}{enegetic particle}
\newacronym{mhd}{MHD}{magnetohydrodynamics}
\newacronym{pic}{PIC}{particle-in-cell}
\newacronym{iter}{ITER}{International Thermonuclear Experimental Reactor}
\newacronym{cfl}{CFL}{Courant–Friedrichs–Lewy}
\newacronym{rk4}{RK4}{4th order Runge-Kutta}
\newacronym{cpu}{CPU}{central processing unit}
\newacronym{gpu}{GPU}{graphics processing unit}
\newacronym{mpi}{MPI}{Message Passing Interface}
\newacronym{shm}{SHM}{MPI Shared Memory}
\newacronym{tae}{TAE}{toroidal Alfvén eigenmode}
\newacronym{rsae}{RSAE}{reversed shear Alfvén eigenmode}
\newacronym{gae}{GAE}{global Alfvén eigenmode}
\newacronym{cae}{CAE}{compressional Alfvén eigenmode}
\newacronym{flr}{FLR}{finite Larmor radius}
\newacronym{zlr}{ZLR}{zero Larmor radius}
\newacronym{fow}{FOW}{finite orbit width}
\newacronym{gs}{G-S}{Grad-Shafranov}

\begin{document}

\begin{frontmatter}

\title{Hybrid simulation of energetic particles interacting with magnetohydrodynamics using a slow manifold algorithm and GPU acceleration}
\author{Chang Liu\corref{mycorrespondingauthor}}
%\email{cliu@pppl.gov}
\cortext[mycorrespondingauthor]{Corresponding author}
\ead{cliu@pppl.gov}
\address{Princeton Plasma Physics Laboratory, Princeton, NJ, 08540, USA}

\author{Stephen C. Jardin}
\address{Princeton Plasma Physics Laboratory, Princeton, NJ, 08540, USA}
\author{Hong Qin}
\address{Princeton Plasma Physics Laboratory, Princeton, NJ, 08540, USA}
\author{Jianyuan Xiao}
\address{University of Science and Technology of China, Hefei, 230026, China}
\author{Nathaniel M. Ferraro}
\address{Princeton Plasma Physics Laboratory, Princeton, NJ, 08540, USA}
\author{Joshua Breslau}
\address{Princeton Plasma Physics Laboratory, Princeton, NJ, 08540, USA}

\begin{abstract}
The hybrid method combining particle-in-cell and magnetohydrodynamics can be used to study the interaction between energetic particles and global plasma modes. In this paper we introduce the M3D-C1-K code, which is developed based on the M3D-C1 finite element code solving the magnetohydrodynamics equations, with a newly developed kinetic module simulating energetic particles. The particle pushing is done using a new algorithm by applying the Boris pusher to the classical Pauli particles to simulate the slow-manifold of particle orbits, with long-term accuracy and fidelity. The particle pushing can be accelerated using GPUs with a significant speedup. The moments of the particles are calculated using the $\delta f$ method, and are coupled into the  magnetohydrodynamics simulation through pressure or current coupling schemes. Several linear simulations of magnetohydrodynamics modes driven by energetic particles have been conducted using M3D-C1-K, including fishbone, toroidal Alfvén eigenmodes and reversed shear Alfvén eigenmodes. Good agreement with previous results from other eigenvalue, kinetic and hybrid codes has been achieved. 
\end{abstract}


\end{frontmatter}

%\maketitle

%\linenumbers

\section{Introfduction}

The physics of \glspl{ep} is an important area of plasma physics and the their confinement is critical to the success of \gls{iter} and future fusion reactors. \glspl{ep} can interact with the bulk plasma and drive \gls{mhd} instabilities, which can cause significant transport of \glspl{ep}. These physics problems must be simulated comprehensively as there are strong kinetic effects associated with \glspl{ep}. A widely used strategy to study \glspl{ep} is the hybrid simulation, which combines the \gls{pic} and the \gls{mhd} simulations. In this method, \glspl{ep} are described with markers carrying density and momentum, and are pushed following the equation of motion of the \glspl{ep} with the electromagnetic fields from the \gls{mhd} simulations. The moments of \glspl{ep} are calculated using the obtained distribution function, where the $\delta f$ method can be used to reduce the noise. The moments are then coupled into the \gls{mhd} equations, which characterizes the energy and momentum exchange between the \glspl{ep} and the bulk plasmas. With such a coupling scheme in the simulation, when the motion of \glspl{ep} is in resonance with some \gls{mhd} modes, the \gls{ep} distribution can be significantly altered near the resonance region and can give strong feedback to the modes. Compared to fully kinetic simulations in which both the \glspl{ep} and the bulk plasmas are described using particles, the hybrid approach can save substantial  simulation time while still keeping the essential physics related to \gls{ep}-\gls{mhd} interaction.

In most of the previously developed hybrid simulation codes\cite{todo_linear_1998,fu_global_2006,kim_impact_2008}, particle pushing is done following the guiding center equations of motion in order to reduce the particle phase space dimension and allow the usage of timesteps larger than the gyro period. It has been observed\cite{qin_variational_2008} that advancing guiding center equations using explicit integration methods like the Runge-Kutta method can lead to breakdown of energy and momentum conservation and large deviation of particle orbits for long time simulations due to the accumulation of numerical error. Recently, a series of methods for pushing the slow manifold of magnetized particles have been developed\cite{xiao_slow_2021}. In these methods, the mirror force is treated as an additional conservative force, which enables us to use full orbit particle pushing algorithms like the Boris algorithm with timesteps larger than the gyro period while still keeping the simplicity and the structure preserving property of the algorithm.

In order to perform long time hybrid simulations to study the physics of \glspl{ep}, we have developed a new hybrid code M3D-C1-K, in which we have implemented one of the slow manifold algorithms introduced in \cite{xiao_slow_2021}, whose essence is to use the Boris algorithm to push the slow manifold of classical Pauli particle orbits. The code is based on the M3D-C1 code\cite{ferraro_calculations_2009}, which solves the \gls{mhd} equations as an initial value problem using high order 3D finite elements. The code can do both linear and nonlinear simulations, and the \gls{mhd} equations can be integrated using fully implicit or semi-implicit methods\cite{jardin_multiple_2012}. The particle pushing is developed with particle based parallelization, and can run on \glspl{gpu} with significant speedup compared to running on \glspl{cpu}. In addition to particle pushing, M3D-C1-K also includes the calculation of the particle distribution function evolution using the $\delta f$ method, and the particle weight is used to calculate the perturbed moments. The moments of the particle distribution function are coupled with the \gls{mhd} equations using one of two schemes, pressure coupling or current coupling, which utilize different orders of moments but are physically equivalent. This new code has been tested with a number of linear simulation problems including the excitation of Alfvén eigenmodes and fishbone modes, and the results agree well with those of other codes.

This paper is organized as follows: in \cref{sec:slow-manifold} we introduce the new slow manifold Boris algorithm used in this code, including a test run showing its conservation property. In \cref{sec:delta-f-method} we introduce the $\delta f$ method and how we calculate the particle weights that are used for deposition. In \cref{sec:coupling-mhd} we discuss the pressure coupling and current coupling schemes and how they are implemented in M3D-C1-K. In \cref{sec:gpu-acceleration} we show how the code utilizes \glspl{gpu} to realize particle based parallelization, and how the data is transferred between \glspl{cpu} and \glspl{gpu}. We also present a comparison of the particle pushing code running on \glspl{cpu} and \glspl{gpu}. In \cref{sec:simulation-result}, we show a series of simulation results using this new code, and a comparison with results from other codes. In \cref{sec:conclusion} we conclude.

\section{Particle pushing with slow manifold algorithm}
\label{sec:slow-manifold}

In M3D-C1-K, a hybrid model is utilized to simulate the physics of the bulk plasma and the \glspl{ep}. The bulk plasma is described by the \gls{mhd} equations which are solved using the finite element method. \glspl{ep} are represented by markers and advanced using the particles' equations of motion, which are calculated using the electromagnetic field information obtained from the \gls{mhd} equations. Then the  \gls{ep} information is coupled back into the \gls{mhd} equations by depositing moment information onto the finite element mesh. This is similar to a \gls{pic} simulation. The difference between this and fully kinetic or gyrokinetic \gls{pic} simulation is that in a fully kinetic simulation, particle density and current are used in the Poisson’s equation and the Ampere’s law to calculate the electromagnetic fields. But in a hybrid model we use the pressure or current from the \glspl{ep} and insert them into the \gls{mhd} equations.

In previously developed hybrid codes like M3D-K\cite{fu_global_2006} and NIMROD\cite{kim_impact_2008}, the orbits of marker particles follow the drift or gyro kinetic equations. For example, the particles' equations of motion implemented in M3D-K can be written as
\begin{equation}
\label{eq:rk4-1}
	\frac{d\mathbf{X}}{dt}=\frac{1}{B^\star}\left[v_\parallel \mathbf{B}^\star-\mathbf{b}\times \left(\mathbf{E}-\frac{\mu}{q}\nabla B\right)\right],
\end{equation}
\begin{equation}
\label{eq:rk4-2}
	m \frac{d v_\parallel}{dt}=\frac{1}{B^\star}\mathbf{B}^\star\cdot \left(q\mathbf{E}-\mu\nabla B\right),
\end{equation}
where
\begin{equation}
	\mathbf{B}^\star=\mathbf{B}+\frac{m v_\parallel}{q}\nabla\times\mathbf{b},
\end{equation}
\begin{equation}
\label{eq:rk4-4}
	B^\star=\mathbf{B}^\star\cdot \mathbf{b}.
\end{equation}
Here $\mathbf{E}$ is the electric field, $\mathbf{B}$ is the magnetic field, $\mathbf{b}=\mathbf{B}/|B|$ is the unit vector in the direction of $\mathbf{B}$, $\mathbf{X}$ is the guiding center location, $v_\parallel$ is the parallel velocity, $\mu$ is the magnetic moment, and $m$ and $q$ are the mass and charge of particles.

The equations of motion are derived from a Lagrangian written in guiding center coordinates following the variational principle\cite{littlejohn_variational_1983}. We can see that in this model, the gyro phase angle is an ignorable coordinate which reduces the explicit phase space from 6D to 5D. The timestep for calculating the equations of motion can be chosen based on the particles’ drift motion, and can be much larger than the gyro period ($2\pi/\Omega$, $\Omega$ is the particle gyro frequency) and can thus save considerable computation time. 

The guiding center equations of motion can be calculated using an explicit integration method like \gls{rk4}. Although \gls{rk4} minimizes the numerical error at every step, it has been shown that the error can accumulate and lead to nonphysical results in long time simulations\cite{qin_variational_2008}. For example, for a collisionless particle moving in a static magnetic field in tokamak geometry, the toroidal angular momentum ($P_\phi=q\psi+mv_\parallel R B_\phi/B$, $\psi$ is the poloidal field flux and $\phi$ is the toroidal direction) and kinetic energy ($E=(1/2)m v_\parallel^2+\mu B$) will not be conserved if using \gls{rk4} for particle pushing, leading to the particle orbit deviating from its original drift motion surface\cite{qin_variational_2009}. This problem can be more serious for particles with large parallel momentum such as energetic particles generated in fusion reactions or energetic electrons such as runaway electrons. To resolve this problem, symplectic algorithms\cite{qin_variational_2008,ellison_degenerate_2018} and structure-preserving methods\cite{liu_collisionless_2016} have been developed, which were designed to preserve physical Casimir invariants when integrating the equations of motion.

In this regard, in M3D-C1-K, in addition to \gls{rk4} integration of guiding center equations, we have implemented an alternative method for particle pushing, which is a volume-preserving slow manifold Boris algorithm. The Boris algorithm has been widely used for pushing particles in magnetic fields. It has been shown to have excellent long time accuracy\cite{qin_why_2013}. Since it was developed for integration of full orbits of magnetized particles, the timestep is chosen to be much smaller than the gyroperiod. However, it has been shown\cite{xiao_slow_2021} that by introducing a mirror force term which behaves like an effective electric force, one can use the Boris algorithm to calculate the slow manifold of a magnetized particle’s orbit, which is close to the guiding center orbit. The mirror force term will give the effect of the gradient drift, while the curvature drift will be given by the Boris algorithm itself. The algorithm can be described as
\begin{equation}
\label{eq:boris-1}
	\frac{\mathbf{x}_l-\mathbf{x}_{l-1}}{\Delta t}=\mathbf{v}_{l-1/2},
\end{equation}
\begin{equation}
\label{eq:boris-2}
	\frac{m}{q}\frac{\mathbf{v}_{l+1/2}-\mathbf{v}_{l-1/2}}{\Delta t}=\mathbf{E}^\dagger(\mathbf{x}_l)+\frac{\mathbf{v}_{l+1/2}+\mathbf{v}_{l-1/2}}{2}\times\mathbf{B}(\mathbf{x}_l)
\end{equation}
where $\mathbf{E}^\dagger=\mathbf{E}-\mu\nabla B$, and $\mathbf{x}_l$ and $\mathbf{v}_{l-1/2}$ characterize the location and velocity of the slow manifold at $l$ and $l-1/2$ timestep. Though \cref{eq:boris-2} looks like an implicit form with $\mathbf{v}_{l+1/2}$ appearing on both sides, it can be calculated explicitly as shown in \cite{qin_why_2013}.

As discussed in \cite{xiao_slow_2021}, by including the mirror force in $\mathbf{E}^\dagger$, the algorithm can be used to push particles with timesteps larger than $2\pi/\Omega$, as long as particles stay close to the slow manifold\footnote{If not initialized accurately, markers may just jump back and forth across the slow manifold which leads to large errors.}. Note that in the Boris algorithm, $\mathbf{x}$ and $\mathbf{v}$ lie on different times with a difference of $1/2 \Delta t$, which is like the leapfrog integration method. To bootstrap the Boris algorithm at the initial timestep, we use \gls{rk4} to advance the guiding center equations of motion (Eqs.~(\ref{eq:rk4-1}-\ref{eq:rk4-4}) from $\mathbf{x}_0$ to $\mathbf{x}_1$, and then use $(\mathbf{x}_1-\mathbf{x}_0)/\Delta t$ as $\mathbf{v}_{1/2}$, to ensures that markers stay close to the slow manifold of particle motion.

\begin{figure}[h]
	\begin{center}
		\includegraphics[width=0.3\linewidth]{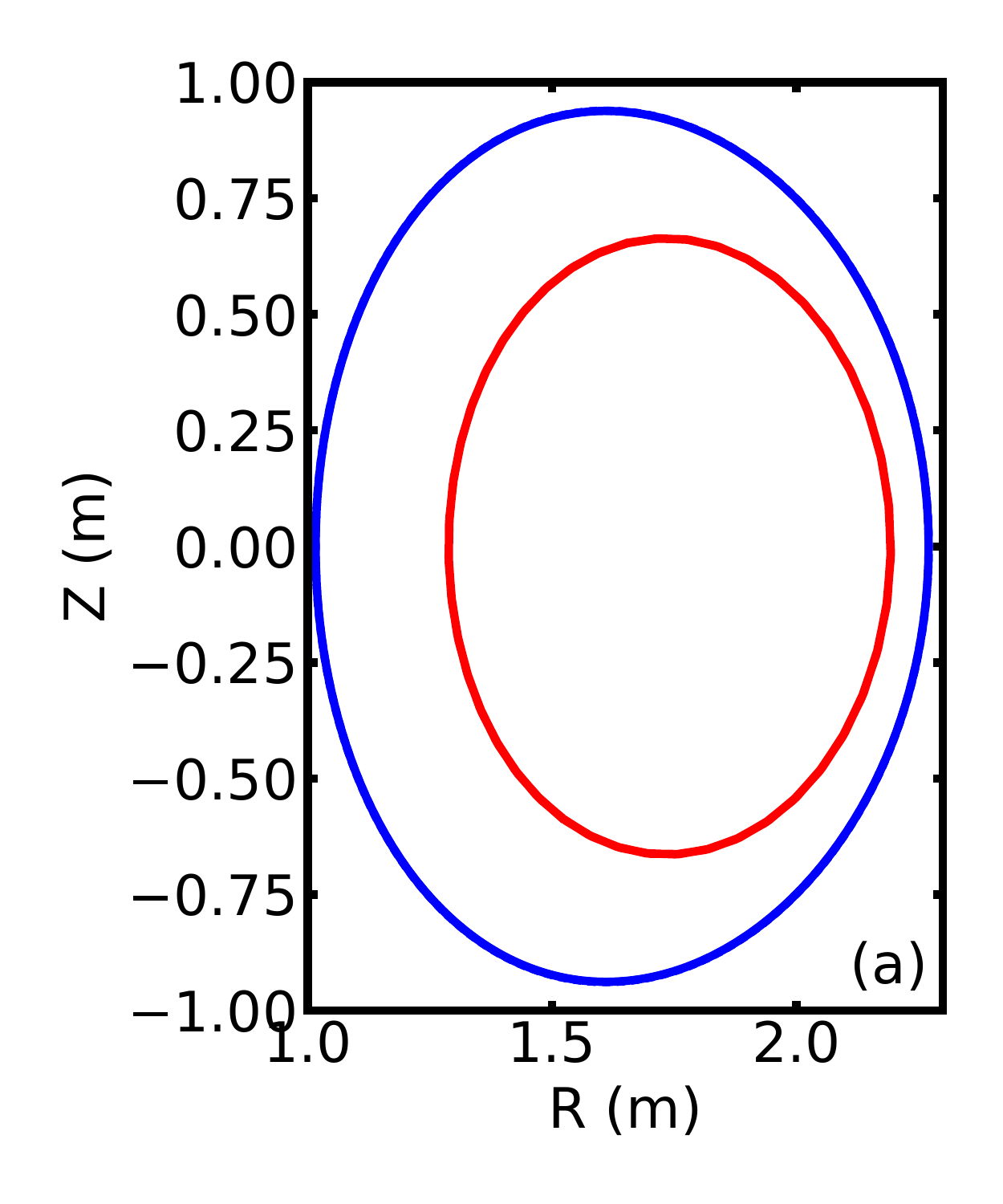}
		\raisebox{0.25\height}{\includegraphics[width=0.3\linewidth]{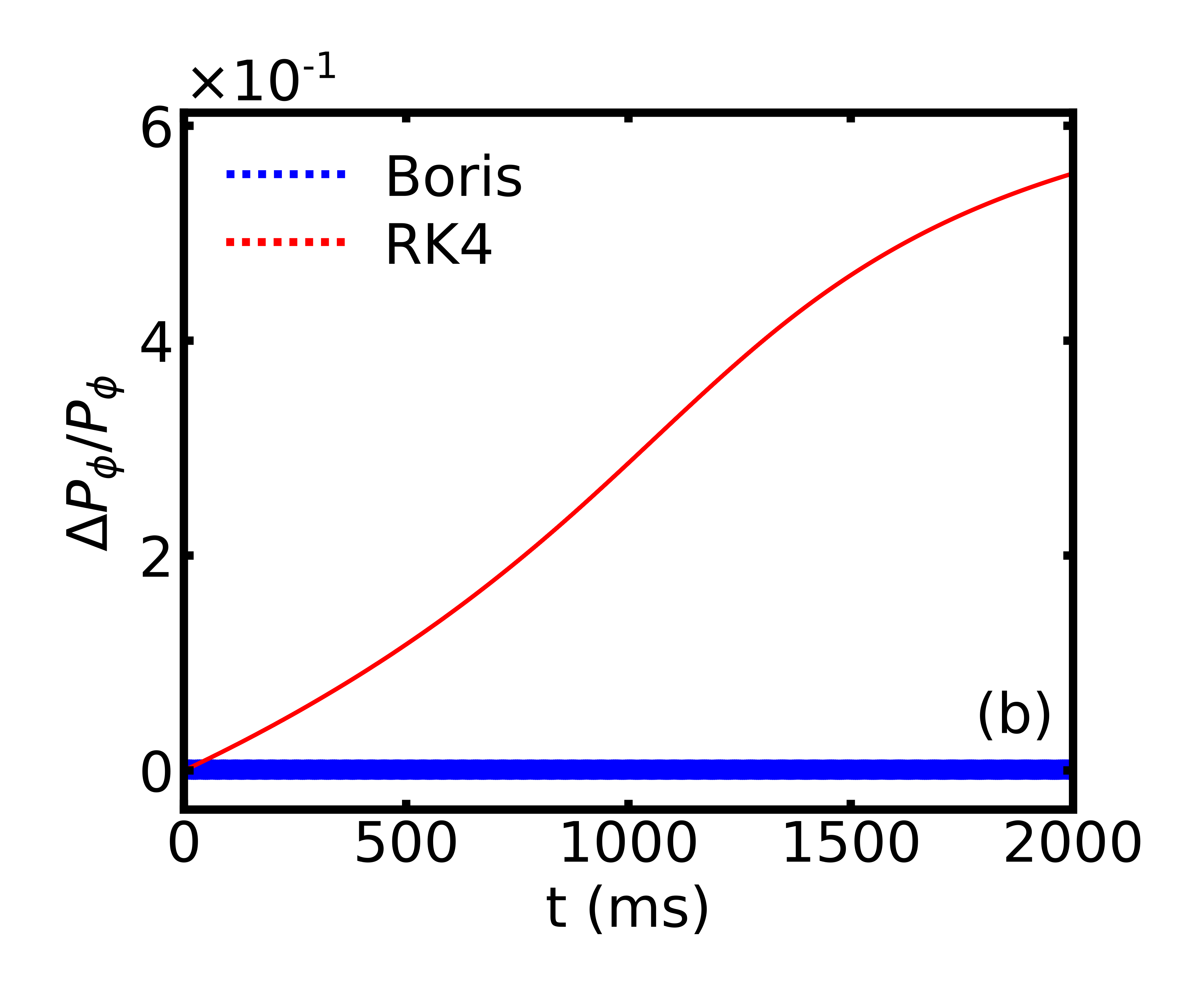}}
		\raisebox{0.25\height}{\includegraphics[width=0.3\linewidth]{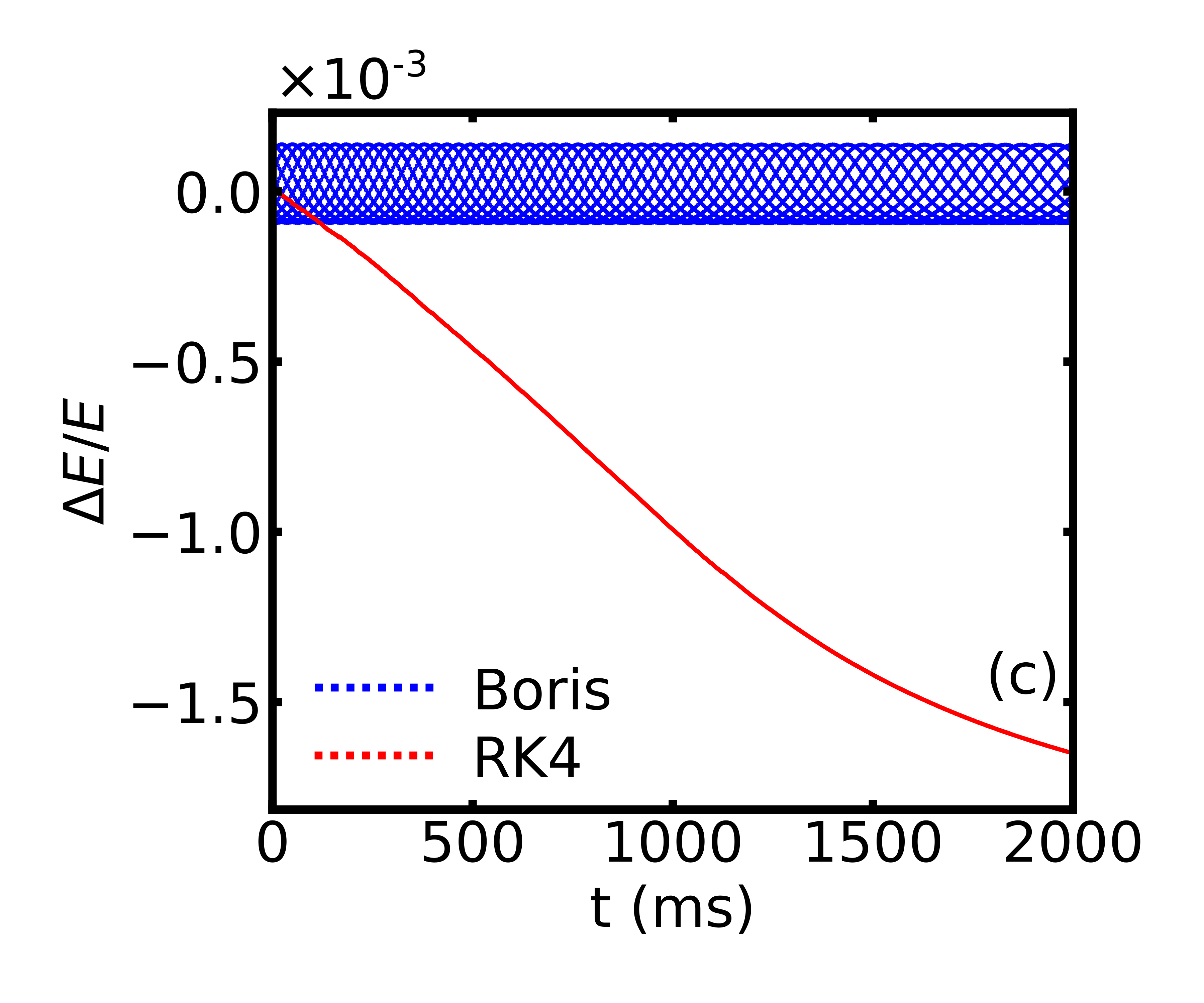}}\\
		\includegraphics[width=0.3\linewidth]{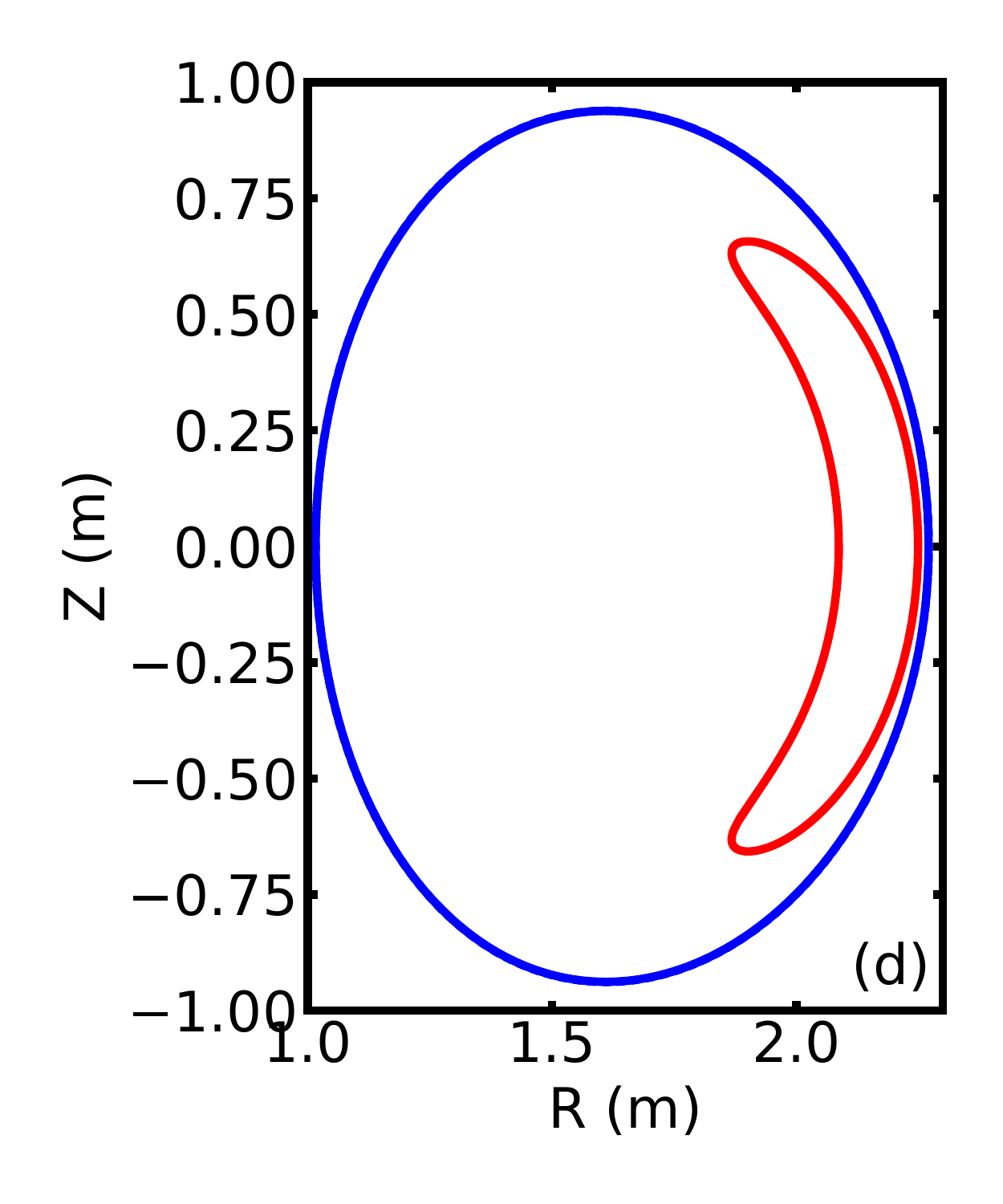}
		\raisebox{0.25\height}{\includegraphics[width=0.3\linewidth]{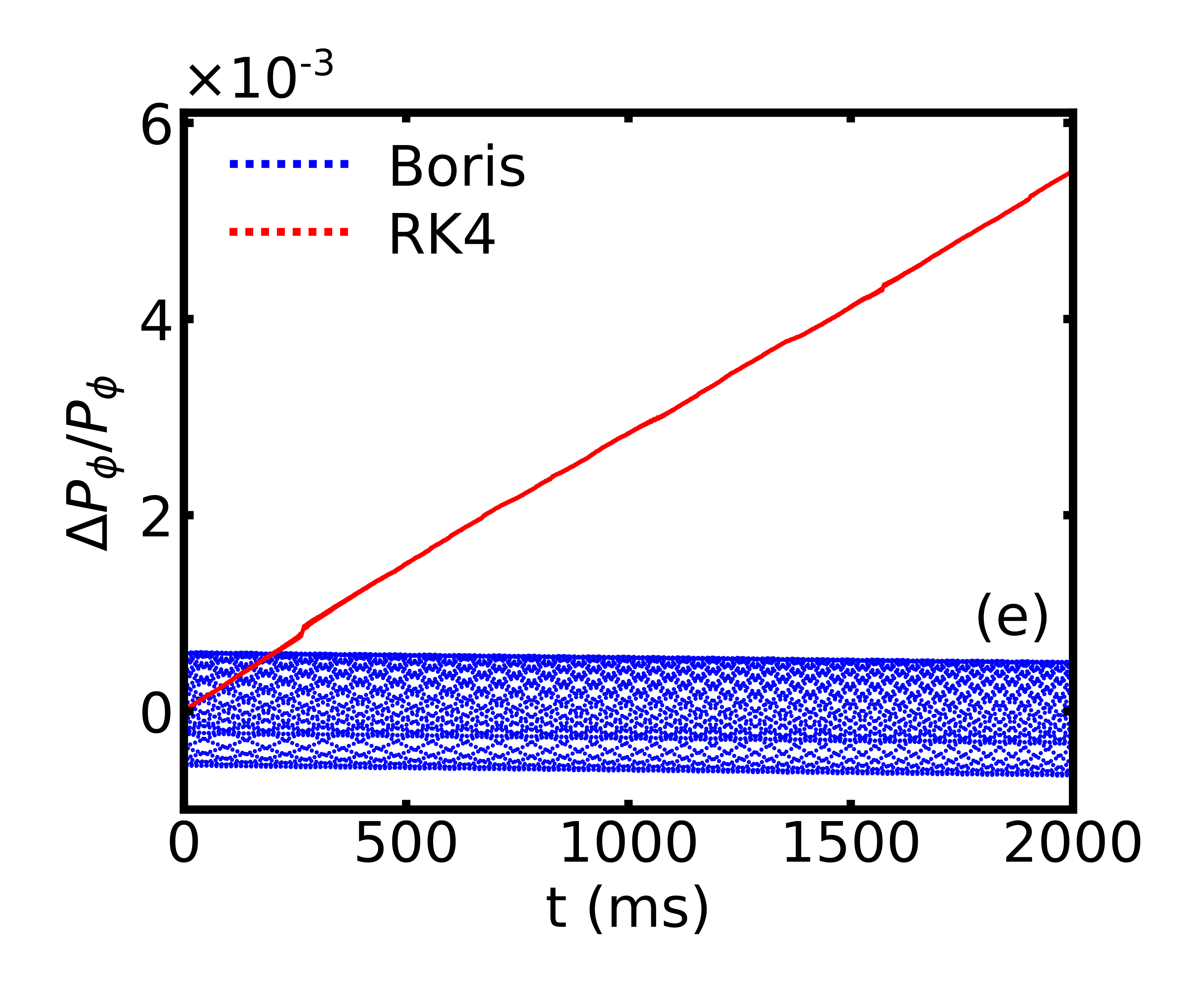}}
		\raisebox{0.25\height}{\includegraphics[width=0.3\linewidth]{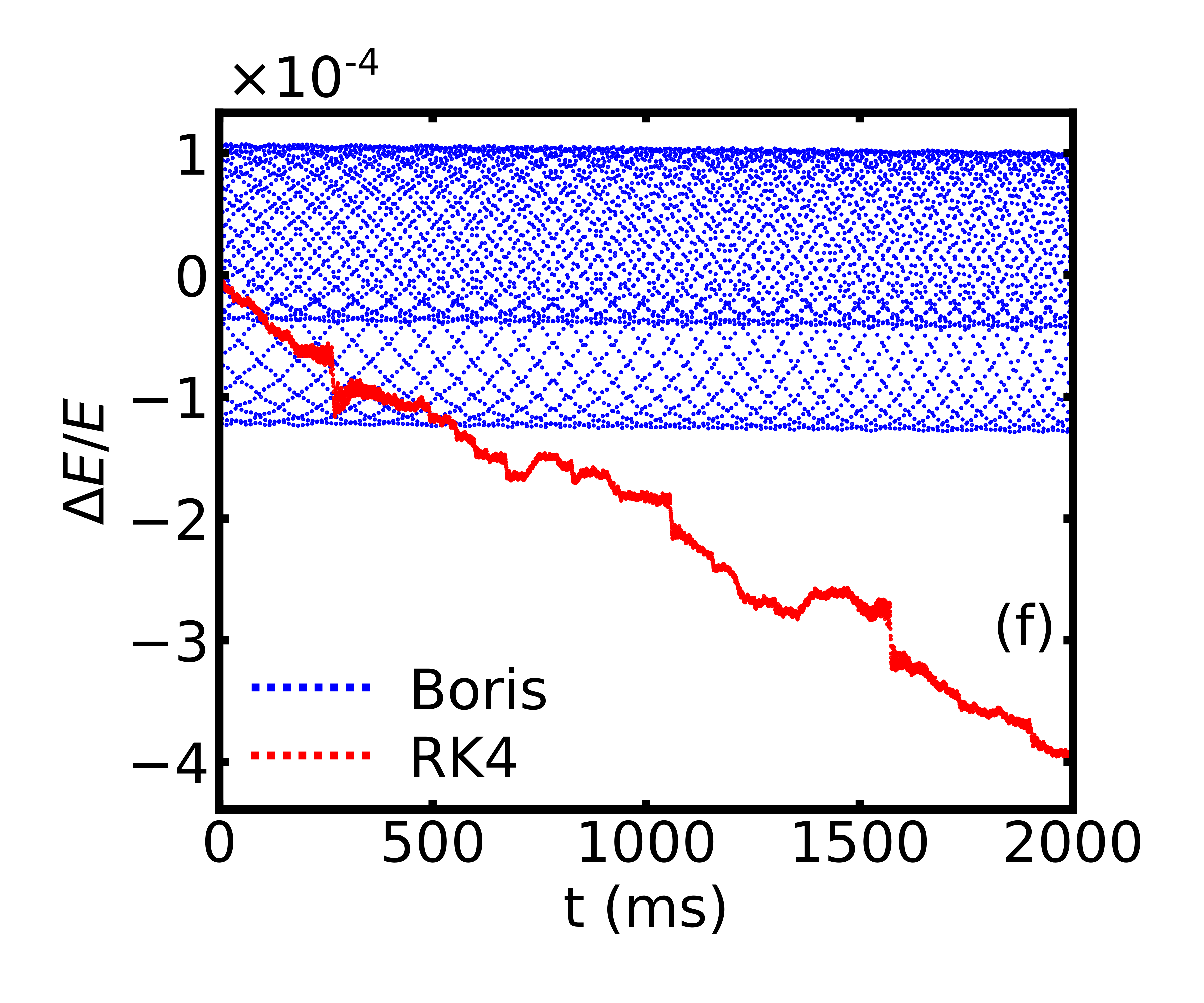}}
	\end{center}
	\caption{\label{fig:boris} The above plots show the simulation results of a passing particle with $v_\parallel=2.4\times 10^6$m/s, $v_\perp=7\times10^5$m/s, including the particle orbit (red line in (a)), relative error of toroidal angular momentum $P_\phi$ (b) and energy $E$ (c) using Boris and \gls{rk4} methods. The below plots show the simulation results of a trapped particle with $v_\parallel=7\times 10^5$m/s, $v_\perp=2.4\times10^6$m/s, including the particle orbit (red line in (d)), relative error of $P_\phi$ (e) and $E$ (f).  }
\end{figure}

To check the conservation property of the slow manifold Boris algorithm, we did a test run to push particles in a static magnetic field without an electric field. The simulation is set up in a DIII-D tokamak like geometry, with minor radius $a=0.67$m, major radius $R=1.67$m, and on-axis magnetic field $B=2$T. Two particles were tested. One is a passing particle with $v_\parallel=2.4\times 10^6$m/s and $v_\perp=7\times10^5$m/s. The other is a trapped particle with $v_\parallel=7\times 10^5$m/s and $v_\perp=2.4\times10^6$m/s. Both are initialized at the low field side. For the integration of the guiding center equations with \gls{rk4} we use a timestep $\Delta t=3.2\times 10^{-7}$s$ \approx 5/(2\pi/\Omega)$, and for the slow manifold Boris algorithm we use a smaller timestep $\Delta t'=1/4 \Delta t$ which gives a similar total computation time as \gls{rk4}. \cref{fig:boris} shows the error of the particles' toroidal angular momentum $P_\phi$ and energy $E$ using the two methods.  We can see that the numerical error of \gls{rk4} will accumulate and reach a significant level for long time simulation, while the error of the Boris algorithm is always bounded. The Boris algorithm shows a better long time conservation property for both $P_\phi$ and $E$, especially for passing particles with large $v_\parallel$, though the benefit is only significant for long time simulations ($t>500$ms). For short time simulations, the error of \gls{rk4} is smaller as it is derived from a higher order integration method.

In addition, the slow manifold Boris algorithm can give a speedup relative to \gls{rk4} when used in M3D-C1-K. In M3D-C1-K, the electromagnetic fields are represented using scalar and vector potentials ($\phi, \mathbf{A}$). When evaluating the fields ($\mathbf{E},\mathbf{B}$) at a specific point during particle pushing, the derivatives of the polynomials are needed. Thus if calculating terms like the magnetic field curvature term in the guiding center equations, one needs to calculate the second order derivatives of the polynomials, which can be time-consuming when using a 3D mesh. After profiling the particle pushing code using \gls{rk4}, it was found that most of the time is spent in the evaluation of the second order derivatives of polynomials. When using the Boris algorithm, the magnetic field curvature term is not needed, and the gradient term $\nabla B$ can be easily calculated by treating $B$ as an additional scalar field, thus only the first order derivative of the polynomials is needed. In addition, although the Boris method can have larger error at each step, it is acceptable since the error will not accumulate. The speedup brought by the Boris algorithm is illustrated in a simulation test in \cref{sec:gpu-acceleration}.

\section{$\delta f$ method and particle weight calculation}
\label{sec:delta-f-method}

The moments of kinetic particles are calculated using their distribution function. In order to reduce the numerical noise, we use the $\delta f$ method to calculate the change of the particle distribution function, meaning that for each marker, in addition to its coordinates, we also need to calculate the evolution of the value of $\delta f=f-f_0$ or particle weight $w=\delta f/f$ during the particle pushing. Here $f_0$ is the equilibrium particle distribution function. The $\delta f$ method can be applied to linear simulations, or nonlinear simulations if the perturbed quantities are not far from their equilibrium values. This is the case, for example, in the Alfvén wave frequency chirping simulations. However, for nonlinear simulations with significant change of quantities, there is no benefits to using this and a full-f method should be used instead.

The evolution of $\delta f$ can be written as
\begin{equation}
	\label{eq:ddeltafdt}
	\frac{d\delta f}{dt}=-\frac{df_0}{dt},
\end{equation}
which is derived from the particle Vlasov equation $df/dt=0$. \cref{eq:ddeltafdt} can also be written as the evolution of $w$ as
\begin{equation}
	\label{eq:dwdt}
	\frac{dw}{dt}=-(1-w)\frac{1}{f_0}\frac{df_0}{dt}.
\end{equation}
In the particle simulation the $dw/dt$ term represents the change of particle weight following its trajectory, and can be calculated during particle pushing. When doing linear simulations, the particle trajectory is calculated using the equilibrium field only. In addition, the $(1-w)$ term in \cref{eq:dwdt} will be replaced by 1, so that \cref{eq:dwdt} only includes linear terms. For nonlinear simulations, the particle trajectory calculation includes both the equilibrium and the perturbed fields.

In the above equations, $df_0/dt$ represents the change of the equilibrium distribution by the perturbed fields, since $df_0/dt=0$ with the equilibrium fields only. Given that there is no electric field in the equilibrium,  $P_\phi$, $E$ and $\mu$ are constants of motion in the absence of perturbations. The time derivative of $f_0$ can then be written as
\begin{equation}
	\frac{df_0}{dt}=\frac{d P_\phi}{dt}\frac{\partial f_0}{\partial P_\phi}+\frac{d E}{dt}\frac{\partial f_0}{\partial E},
\end{equation}
and $\partial f_0/\partial P_\phi$ and $\partial f_0/\partial E$ can be calculated from the analytical expression of $f_0$ using the chain rule. Here we assume $\mu$ will not change with perturbed fields following the approximation of guiding center. Since $P_\phi$ and $E$ will only be changed by the perturbed fields, their time derivatives can be expressed as
\begin{equation}
	\frac{dP_\phi}{dt}=\left(\frac{d\mathbf{X}}{dt}\right)_1\cdot\nabla\psi+\left(\frac{d v_\parallel}{dt}\right)_1 mRB_\phi/B,
\end{equation}
\begin{equation}
	\frac{dE}{dt}=q\mathbf{v}\cdot\mathbf{E}_1+\mu\frac{\partial B_{1 \parallel}}{\partial t}.
\end{equation}
In a linear simulation, the $\left(\dots\right)_1$ terms can be expressed as
\begin{equation}
	\left(\frac{d\mathbf{X}}{dt}\right)_1=\frac{\mathbf{E}_1\times\mathbf{B}_0}{B_0^2}+v_\parallel\frac{\mathbf{B}_1}{B_0},
\end{equation}
\begin{equation}
	\left(\frac{d v_\parallel}{dt}\right)_1=q\mathbf{E}\cdot\mathbf{B}/B-\mathbf{b}_0\cdot\mu\nabla B_{1 \parallel},
\end{equation}
where $\mathbf{E}_1$ and $\mathbf{B}_1$ are the perturbed electric and magnetic fields, and $B_{1,\parallel}=\mathbf{b}_0\cdot\mathbf{B}_1$. For a nonlinear simulation, $\left(d\mathbf{X}/dt\right)_1$ and $\left(dv_\parallel/dt\right)_1$ can be obtained by calculating the difference between $d\mathbf{X}/dt$ and $dv_\parallel/dt$ from the Boris algorithm including all the perturbed fields, with $\left(d\mathbf{X}/dt\right)_0$ and $(dv_\parallel/dt)_0$ using only the equilibrium fields following the guiding center equation, in order to include all the nonlinear contributions. $\partial B_{1\parallel}/\partial t$ and $\nabla B_{1\parallel}$ are calculated similarly by taking the difference of the results with and without perturbed fields.

To include the \gls{flr} effect related to physics on small spatial scales comparable to the gyroradius, one can use orbit-averaged fields $\langle\mathbf{E}_1\rangle$, $\langle\mathbf{B}_1\rangle$ in the above equations to replace the fields $\mathbf{E}_1$, $\mathbf{B}_1$ evaluated at the guiding center, like
\begin{align}
	\label{eq:gyroaverage}
	\langle\mathbf{B}_1\rangle(\mathbf{X})&=\int \mathbf{B}_1(x)\delta(\mathbf{x}-\mathbf{X}-\bm{\rho}_L)d\mathbf{x} d\theta,\nonumber\\
	&\approx \frac{1}{4}\sum_{j=1}^4 \mathbf{B}_1\left(\mathbf{X}+\bm{\rho}_{L,j}\right).
\end{align}
Here $\bm{\rho}_L=\mathbf{v}_\perp\times\mathbf{b}/\Omega$ is the gyro radius vector, $\mathbf{v}_\perp$ is the particle velocity perpendicular to the magnetic field calculated from $\mu$, and $\theta$ is the gyro phase angle. The integration can be approximately calculated using the 4-point averaging scheme\cite{lee_gyrokinetic_1987,wang_gyro-kinetic_2006}, where $\bm{\rho}_{L,j}$ are 4 vectors with length $|\rho_L|$ and are uniformly distributed in $\theta$.

The calculation of the change in particle weights needs particle's $\mathbf{v}$ and $\mathbf{x}$ (required for field evaluation) at the same time. When pushing particles using the Boris algorithm, we take $\mathbf{v}_l=\left(\mathbf{v}_{l-1/2}+\mathbf{v}_{l+1/2}\right)/2$, and use $\mathbf{x}_l$ and $\mathbf{v}_l$ in the integration of the weight equation.

After obtaining $\delta f$ or $w$, the moments can be calculated from them. The parallel and perpendicular pressure can be calculated as
\begin{align}
	\label{eq:ppar}
	\delta P_\parallel(\mathbf{x})&=\int mv_\parallel^2 w f B^*dv_\parallel d\mu d\theta,\nonumber\\
	&\approx \sum_k mv_\parallel^2 w_k \frac{f_k}{g_k}B^* S\left(\mathbf{x}-\mathbf{x}_k\right),
\end{align}
\begin{align}
	\label{eq:pperp}
	\delta P_\perp(\mathbf{x})&=\int \mu B \left(w+\frac{B_{1 \parallel}}{B_0}\right) f B^*dv_\parallel d\mu d\theta,\nonumber\\
	&=\sum_k \mu B \left(w_k+\frac{B_{1\parallel}}{B_0}\right) \frac{f_k}{g_k}B^* S\left(\mathbf{x}-\mathbf{x}_k\right).
\end{align}
Here $\sum_k$ is the summation of all the particle markers, $B^*$ characterizes the phase space volume and is used as the Jacobian for the phase space integral, and $S$ is the shape function used for particle deposition. In the calculation of $\delta P_\perp$ the change of perpendicular pressure due to the variation of $B_\parallel$ is taken into account. Here $g$ represents the distribution of loaded makers, which depends on how the markers are initialized. If the makers are initialized uniformly in phase space, then $g=B^*$ and the summation should include $f$ in the summation, or include an additional $f/g$ term in the weight evolution equation like in M3D-K\cite{fu_global_2006}. In M3D-C1-K, we initialize the markers following the same distribution function $f_0$ using the Monte Carlo method in order to reduce the total number of markers while keeping a low noise level. With this implementation, the marker distribution will then follow the evolution of $B^* f$ during the simulation, and the $f/g$ term in the summations of Eqs.~(\ref{eq:ppar}) and (\ref{eq:pperp}) can be ignored.

Note that according to Eqs.~(\ref{eq:ppar}) and (\ref{eq:pperp}) the change of integration Jacobian $B^*$ can also affect the particle moments. For example, when affected by a compressing magnetic field ($\nabla\times\mathbf{E}\ne 0$), the particle distribution which is initially homogeneous in space and energy can be compressed by the $\mathbf{E}\times\mathbf{B}$ velocities and form a gradient. This effect can be captured by the particle pushing since it is equivalent to solving a continuity equation as pointed out by \cite{lee_gyrokinetic_1987}, so that $g=B^* f$ can be kept. However, it cannot be captured by the $df_0/dt$ term as there is no gradient in the initial particle distribution function. To address this issue, we can follow the discussion in \cite{belova_hybrid_1997} and use $d=w+(1-w)B^*_1/B^*$ to replace $w$ in the summation in Eqs.~(\ref{eq:ppar}) and (\ref{eq:pperp}), which was also used in the M3D-K implementation. For a linear simulation, the definition of $d$ is  $d=w+B^*_1/B^*$ which only keeps the linear terms. Note that with this additional term and the $B_{1 \parallel}/B_0$ term in \cref{eq:pperp}, \glspl{ep} can behave like plasma with heat capacity ratio $\gamma=2$ in the perpendicular direction.

In a finite-element representation, the summations in Eqs.~(\ref{eq:ppar}) and (\ref{eq:pperp}) can be calculated using the Galerkin method to obtain the particle pressure fields, by multiplying with a test function $\nu_i$ and integrate in the elements. This can be written as
\begin{equation}
	\label{eq:ppar-galerkin}
	\int \nu_i \delta P_\parallel J d\mathbf{x}=\sum_k w_k mv_{k,\parallel}^2 \int \nu_i(\mathbf{x}) J(\mathbf{x}) S(\mathbf{x}-\mathbf{x}_k) d\mathbf{x},
\end{equation}
\begin{equation}
	\label{eq:pper-galerkin}
	\int \nu_i \delta P_\perp J d\mathbf{x}=\sum_k w_k \mu_k B(\mathbf{x}_k) \int \nu_i(\mathbf{x}) J(\mathbf{x}) S(\mathbf{x}-\mathbf{x}_k) d\mathbf{x}.
\end{equation}
The polynomial coefficients can be obtained by solving the mass matrix. If we take $S$ as a $\delta-$function, the integral can be reduced and the whole calculation is significantly simplified.
However, since in M3D-C1 high order polynomials are used for the test functions, the obtained pressure fields can be spiky. One can use a different $S$ like a tent function with a finite width to get a smoother result, but this means that we also need to use a finite-width shape function when evaluating the field at the particle’s location to make the whole scheme self-consistent, which can complicate the particle pushing and slow down the computation. For the linear simulations discussed in \cref{sec:simulation-result}, we use $\delta-$function as the particle shape function.

When performing simulations including the \gls{flr} effect, the pressure deposition should also be changed following the orbit average scheme with 4-point averaging. The $S\left(\mathbf{x}-\mathbf{x}_k\right)$ terms in Eqs.~(\ref{eq:ppar-galerkin}) and (\ref{eq:pper-galerkin}) should be replaced by $1/4 \sum_{j=1}^4 S\left(\mathbf{x}-\mathbf{X}_k-\bm{\rho}_j\right)$, which means that each particle will contribute to pressure deposition at 4 points along its gyro orbit. This implementation is consistent with the field evaluation in \cref{eq:gyroaverage}.

\section{Coupling to MHD equations}
\label{sec:coupling-mhd}

In the calculation of the contribution of \glspl{ep} to the \gls{mhd} equations, we assume that the density of energetic particles ($n_h$) is small compared to the bulk ion density ($n$). In this case, the major contribution of \glspl{ep} lies in the \gls{mhd} momentum equation. Following different assumptions on the meaning of the \gls{mhd} momentum equation, one can use either pressure coupling or current coupling schemes to represent this contribution.

If we assume that the \gls{mhd} momentum equation describes the change of total momentum including both the energetic particles and the rest of the ions and electrons (bulk plasma), the terms related to the \gls{ep} momentum change and forces should be included. In that case, the \gls{mhd} momentum equation can be written as
\begin{equation}
	\rho\left(\frac{\partial\mathbf{V}}{\partial t}\right)+\rho(\mathbf{V}\cdot\nabla\mathbf{V})+\frac{\partial \mathbf{K}_h}{\partial t}=\mathbf{J}\times\mathbf{B}-\nabla p-\nabla\cdot\mathbf{P}_{h},
\end{equation}
where $\rho$ is the bulk plasma density, $\mathbf{V}$ is the bulk plasma velocity. $\mathbf{J}=\nabla\times\mathbf{B}$ is the total current, $p$ is the bulk plasma pressure, and $\mathbf{P}_h=P_\parallel \mathbf{b}\mathbf{b}+P_\perp \left(\mathbf{I}-\mathbf{b}\mathbf{b}\right)$ is the total \gls{ep} pressure tensor.
To use the result of the $\delta f$ method, one can subtract the equilibrium force balance equation 
\begin{equation}
	\mathbf{J}_0\times\mathbf{B}_0=\nabla p_0+\nabla p_{\mathrm{h0}},
\end{equation}
to only calculate the evolution of the perturbed field. Here we assume that the \gls{ep} equilibrium pressure is isotropic. The momentum equation then becomes
\begin{align}
	\label{eq:momentum-subtract}
	\rho\left(\frac{\partial\mathbf{V}}{\partial t}\right)+\rho(\mathbf{V}\cdot\nabla\mathbf{V})+\frac{\partial \mathbf{K}_h}{\partial t}=&\mathbf{J}_0\times\mathbf{B}_1+\mathbf{J}_1\times\mathbf{B}_0+\mathbf{J}_1\times\mathbf{B}_1\nonumber\\
	&-\nabla \delta p-\nabla\cdot\delta\mathbf{P}_{hot},
\end{align}
where $\mathbf{J}_1=\nabla\times\mathbf{B}_1$ and $\delta\mathbf{P}_{hot}$ is calculated from $\delta f$ like in Eqs.~(\ref{eq:ppar}) and (\ref{eq:pperp}). This method is called "pressure coupling" and is implemented in M3D-K\cite{fu_global_2006}. Note that in M3D-K, the $\partial \mathbf{K}_h/\partial t$ term is ignored assuming the \gls{ep} momentum is small compared to the bulk momentum.

If we assume that the \gls{mhd} momentum equation describes the momentum change of bulk plasma only and does not include the \glspl{ep}, then it should be instead written as
\begin{equation}
	\rho\left(\frac{\partial\mathbf{V}}{\partial t}\right)+(\mathbf{V}\cdot\nabla\mathbf{V})=\left(\mathbf{J}-\mathbf{J}_{h}\right)\times\mathbf{B}-\nabla p
\end{equation}
where $\mathbf{J}_h$ is the \gls{ep} current, and $\mathbf{J}-\mathbf{J}_h$ is the current from the bulk plasma. Here \gls{ep} is coupled into the \gls{mhd} equation through $\mathbf{J}_h$ rather than $\mathbf{P}_h$, therefore this method is called “current coupling”. Note that in this equation we do not include the electric force on the bulk plasma $-q n_h \mathbf{E} $, which was present in the current coupling scheme in \cite{park_threedimensional_1992,park_plasma_1999} due to the fact that the bulk plasma is non-neutral. The reason is that this term will cancel the $\mathbf{J}_h\times\mathbf{B}$ term with \gls{ep} current due to the $\mathbf{E}\times\mathbf{B}$ drift, since the $\mathbf{E}\times\mathbf{B}$ drift will cause both ions and electron to move at the same velocity with their currents canceling\cite{zhao_simulation_2020}.

$\mathbf{J}_h$ includes currents from the parallel motion ($J_{h,\parallel}$), the current due to the drift motion ($\mathbf{J}_{h,D}$), and the magnetization current which is due to the gyro motion of \glspl{ep} ($\mathbf{J}_{h,M}$). The first two kinds of current can be calculated using the result of $d\mathbf{X}/dt$ from the guiding center equation of motion or the slow manifold Boris method. Note that $J_{h,\parallel}$ will not contribute to the $\mathbf{J}\times\mathbf{B}$ force in the momentum equation. $\mathbf{J}_{h,M}$ should be calculated following a pull-back transformation\cite{qin_pullback_2004}, and the result can be written as
\begin{align}
	\mathbf{J}_{h,M}(x)&=\int \dot{\bm{\rho}}\delta(\mathbf{X}+\bm{\rho}-\mathbf{x})f B^*d^3\mathbf{X}dv_\parallel d\mu d\theta,\nonumber\\
	&=\nabla\times \mathbf{M},
\end{align}
where $\mathbf{M}=P_\perp \mathbf{b}/B$. If we take the drift kinetic limit and choose a simple representation of drift velocity including the curvature and gradient drifts,
\begin{equation}
	\mathbf{v}_D=\frac{mv_\parallel^2}{qB}\nabla\times\mathbf{b}+\frac{\mu}{qB}\mathbf{b}\times\nabla B,
\end{equation}
then $\mathbf{J}_{h}\times \mathbf{B}$ can be simplified as
\begin{align}
	\label{eq:jhxb}
	\mathbf{J}_{h}\times \mathbf{B}&=\left[q\int \mathbf{v}_D f B^* dv_\parallel d\mu d\theta+\nabla\times \mathbf{M}\right]\times\mathbf{B},\nonumber\\
	&=P_\parallel \mathbf{b}\cdot\nabla\mathbf{b}-P_\perp\nabla\ln B\times\mathbf{b}\times\mathbf{b}-\nabla\times\left(\frac{P_\perp}{B}\mathbf{b}\right)\times\mathbf{B},\nonumber\\
	&=\left[\nabla P_\perp+\nabla\cdot\left[\left(P_\parallel-P_\perp\right) \mathbf{b}\mathbf{b}\right]\right]\times\mathbf{b}\times\mathbf{b}.
\end{align}
which is close to the $\nabla\cdot \mathbf{P}_h$ term in the pressure coupling scheme, except that here the component parallel to $\mathbf{b}$ is eliminated by the $\times\mathbf{b}\times\mathbf{b}$ operator. This means that we can use the result of $\mathbf{P}_\parallel$ and $\mathbf{P}_\perp$ calculated from Eqs.~(\ref{eq:ppar}) and (\ref{eq:pperp}) for both pressure and current coupling schemes, rather than calculating $\mathbf{J}_h$ separately.

When doing a $\delta f$ simulation, one should subtract the equilibrium force balance equation like in \cref{eq:momentum-subtract},
\begin{equation}
	\rho\left(\frac{\partial\mathbf{V}}{\partial t}\right)+\rho(\mathbf{V}\cdot\nabla\mathbf{V})=(\mathbf{J}_0-\mathbf{J}_{h0})\times\mathbf{B}_1+(\mathbf{J}_1-\delta\mathbf{J}_h)\times(\mathbf{B}_0+\mathbf{B}_1)-\nabla \delta p.
\end{equation}
The pressure terms in \cref{eq:jhxb} should be replaced by $\delta P_\parallel$ and $\delta P_\perp$ to give the result of $\delta \mathbf{J}_h\times(\mathbf{B}_0+\mathbf{B}_1)$. Assuming that the equilibrium \gls{ep} current $\mathbf{J}_{h0}$ is perpendicular to $\mathbf{B}_0$ and satisfies the force balance $\mathbf{J}_{h0}\times\mathbf{B}_0=\nabla p_{h0}$, this force of $\mathbf{J}_{h,0}\times\mathbf{B}_1$ can be written as
\begin{equation}
	\mathbf{J}_{h,0}\times\mathbf{B}_1=\mathbf{b}_0\frac{\mathbf{B}_1}{B_0}\cdot \nabla p_{hot}.
\end{equation}

This simplified current coupling scheme was implemented in the  MEGA code\cite{todo_linear_1998}. Note that in the pressure coupling scheme in \cite{park_plasma_1999}, only the perpendicular part of $\nabla\cdot \mathbf{P}_h$ is added in the momentum equation, which is exactly the same as the result in \cref{eq:jhxb} and is equivalent to the simplified current coupling scheme. The reason is that, assuming the perpendicular motion of both bulk plasma and \glspl{ep} are dominated by $\mathbf{E}\times\mathbf{B}$ drifts, then $\mathbf{K}_h$ in the perpendicular direction is much smaller compared to $\rho\mathbf{V}$ as $n_h\ll n$. However, in the parallel direction $\partial \mathbf{K}_h/\partial t$ cannot be safely ignored. In the pressure coupling scheme implemented in M3D-C1-K, we include all the components of the $\nabla\cdot \mathbf{P}_h$ term and ignore the $\partial \mathbf{K}_h/\partial t$ in all directions like in M3D-K. We find that for all the simulations we have conducted, the two coupling schemes give almost the same results.

\section{GPU acceleration of particle pushing}
\label{sec:gpu-acceleration}

The M3D-C1 code was developed using the distributed memory parallelization model with \gls{mpi}. The whole 3D mesh is decomposed into the same number of subdomains as the number of \gls{cpu} processes. Each process is responsible for calculating the elements of the \gls{mhd} equation matrices for one subdomain, and only manages the memory of fields within it. This is called "domain-based parallelization". When developing the particle pushing code for M3D-C1-K, we used the "particle-based parallelization" and "shared memory model" instead. We find that if we stuck with the domain-based model, the code would then need to take care of particles moving from one subdomain to another, which would involve frequent communication between different processes or threads that can significantly slow down the computation. In the particle-based parallelization, each parallel thread takes care of pushing one particle for several timesteps independent of other threads. Therefore this model is suitable for large-scale parallel computing using \glspl{gpu}. This strategy of particle-based parallelization is also used in many gyrokinetic codes like GTC\cite{zhang_heterogeneous_2018} and GTS.

In the development of M3D-C1-K, we utilized \glspl{gpu} to accelerate particle pushing and particle weight calculation, which is the most time-consuming part of the kinetic module. The particle pushing code is developed using OpenACC. OpenACC is a coding standard similar to OpenMP, which provides a list of directives to help write parallel computing code and simplify data communication operations between hosts and accelerator devices (such as \glspl{gpu}). We also implement the multi-thread parallelization of particle pushing on multi-core \glspl{cpu} using OpenMP, so that the code can run on just \glspl{cpu} or with \glspl{gpu} by setting compilation directives. The calculation of the \gls{mhd} equation finite element matrix and the matrix solving is still done by the M3D-C1 code using \glspl{cpu}.

In the implementation of particle-based parallelization, each particle pushing thread must have access to the electromagnetic field information in the whole mesh, so that the particle can move to an arbitrary location in the mesh without performing extra communication. This means that the field information must be collected from each \gls{cpu} processes after the \gls{mhd} calculation and uploaded to the shared memory of each \gls{gpu}. For most modern \glspl{gpu}, the memory is large enough to store the field information of the whole 3D mesh. The data collection on \gls{cpu} processes is done utilizing the \gls{shm} model introduced in \gls{mpi}-3, which can accelerate the communication between processes on the same computation node. For communication between different nodes, the classical message communication interface is used. After the pushing, the particle information needs to be downloaded from \glspl{gpu} and distributed into the distributed memory of each \gls{mpi} processes. The data distribution work and the calculation of $P_\parallel$ and $P_\perp$ for pressure or current coupling is done using \glspl{cpu}.

The fields and particles are evolved separately in M3D-C1-K. The field is evolved according to the \gls{mhd} equations and is integrated using the implicit or semi-implicit method\cite{jardin_multiple_2012}. The implicit \gls{mhd} timestep is not limited by the \gls{cfl} condition and is chosen according to the physical timescale of the problem being studied. The particle pushing and weight calculation are done between the integral of two adjacent \gls{mhd} timesteps. It has subcycles for particle pushing in order to increase the accuracy of particle orbit calculation. The transfer of field information is done before the beginning of the particle pushing subcycles, which can save time for communication between \glspl{cpu} and \glspl{gpu}. During the subcycles, the fields are assumed to be static.

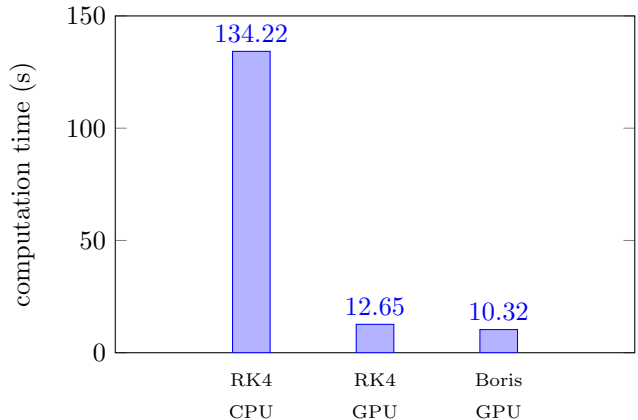
\begin{figure}[h]
	\begin{center}
		\begin{tikzpicture}
			\begin{axis}[
				ybar,
				enlarge x limits=0.55,
				bar width=.5cm,
				width=.7\textwidth,
				height=.5\textwidth,
				symbolic x coords={cpu,gpu,boris},
				xtick=data,
				xtick style={draw=none},
				xticklabel style={align=center,font=\scriptsize},
				xticklabels={RK4\\ CPU, RK4\\ GPU,Boris\\ GPU},
				nodes near coords,
				nodes near coords align={vertical},
				ymin=0,ymax=150,
				ylabel={computation time (s)},
				]
				\addplot coordinates {(cpu,134.22) (gpu,12.65) (boris,10.32)};
			\end{axis}
		\end{tikzpicture}
	\end{center}
	\caption{\label{fig:computation-time}Computation time for pushing 4 million particles for 50 timesteps in a 3D mesh with 4 toroidal planes (5679 elements per plane) using different methods and processors.}
\end{figure}

The performance benchmark of the particle pushing code in M3D-C1-K on \glspl{cpu} and \glspl{gpu} is shown in \cref{fig:computation-time}. After porting the code to \glspl{gpu} without any modification to the algorithm, we get about 11 times speed up when pushing 16 million particles for 50 steps. The benchmark was done on the Summit cluster using four nodes. The \gls{cpu} run utilizes 8 IBM POWER9 \glspl{cpu} with 22 SIMD Multi-Core (SMC) on each processor. The \gls{gpu} run utilizes 24 NVIDIA Tesla V100 \glspl{gpu}. The simulation is set in a 3D mesh with a DIII-D like geometry, with 4 toroidal planes and 5679 elements per plane. The particles are uniformly distributed in the 3D mesh with a Maxwellian distribution. If we use the slow manifold Boris algorithm with timestep $1/4$ of that of \gls{rk4} on \glspl{gpu}, we can get additional speedup due to simplification of the field evaluation as discussed in \cref{sec:slow-manifold}. More speedup can be achieved by further optimization, for example, by improving the coalesce of the \gls{gpu} memory access and using single-precision floating-point arithmetic, but the overall performance will not be significant improved as the computation time for particle pushing in the simulation is already close to the computation time spent on the \gls{cpu} for the \gls{mhd} calculation.

\section{Simulation results}
\label{sec:simulation-result}

In this section we show the linear simulation results of M3D-C1-K, including fishbone, \gls{tae}, and \gls{rsae}. The results are compared with those from other codes, including the mode frequency, growth rate, and structure.

\subsection{Linear fishbone simulation}

For the linear fishbone simulation we followed the setup in \cite{fu_global_2006}, which includes a benchmark study of linear fishbone simulations between M3D-K and NOVA-K in a large aspect ratio circular tokamak. The setup was also used for a benchmark between NIMROD and M3D-K in \cite{kim_impact_2008}. A circular tokamak with $R = 1$m and $a = 0.361925$m was chosen for the test. The plasma consists of hydrogen ions whose density is uniform with $n_0=2.489\times 10^{20}$m$^{-3}$. The total pressure profile is $p(\psi)=p_0 \exp(-\psi/0.25)$, where $\psi$ is the normalized poloidal flux ranging from 0 at the magnetic axis to 1 at the boundary. The central pressure $p_0=16335$Pa and the central total plasma beta $\beta_{total}$ is 8\%. The toroidal field at the magnetic axis is $B_T=1$T. The safety factor ($q$) profile is given by an analytical expression,
\begin{equation}
	q=q_0+\psi\left[q_1-q_0+\left(q'_1-q_1+q_0\right)\frac{\left(1-\psi_s\right)\left(\psi-1\right)}{\psi-\psi_s}\right],
\end{equation}
where $q_0=0.6$ and $q'_0=0.78$ are the value and derivative of $q$ at $\psi=0$, $q_1=2.5$ and $q'_1=5.0$ are the value and derivative of $q$ at $\psi=1$. $\psi_s=\left(q'_1-q_1+q_0\right)/\left(q'_0+q'_1-2q_1+2q_0\right)$.

The density profile of \glspl{ep} has the same shape as the plasma pressure profile. In momentum space it follows an isotropic slowing down distribution given by 
\begin{equation}
	f(v)=\frac{H(v_0-v)}{v^3+v_c^3},
\end{equation}
where $v_0=3.9\times 10^6$m/s is the maximum velocity of \glspl{ep} and $v_c=0.58v_0$ is the critical velocity. The same value of $v_0$ and $v_c$ is used for all flux surfaces. Since the \gls{ep} density and pressure follow the same spatial profile as the plasma pressure, we can vary the value of the \gls{ep} density and the bulk plasma pressure to change the ratio of $\beta_h/\beta_{total}$ ($\beta_h$ is the \gls{ep} pressure beta) while keeping the total pressure profile fixed. Note that when initializing the \gls{ep} distribution we did not consider the average value of $\psi$ for passing and trapped particles like the calculation in \cite{fu_global_2006}. Instead, we just used the local value of $\psi$ for \gls{ep} initialization. In order to satisfy the very small value of normalized Larmor radius used in \cite{fu_global_2006}, $\rho_L=v_0/\left(\Omega a\right)=0.0125$, we use a reduced \gls{ep} ion mass $m_{EP}=0.11 m_H$ ($m_H$ is the hydrogen mass). This can help reduce the \gls{fow} effect of \glspl{ep}.

\begin{figure}[h]
	\begin{center}
		\includegraphics[width=0.39\linewidth]{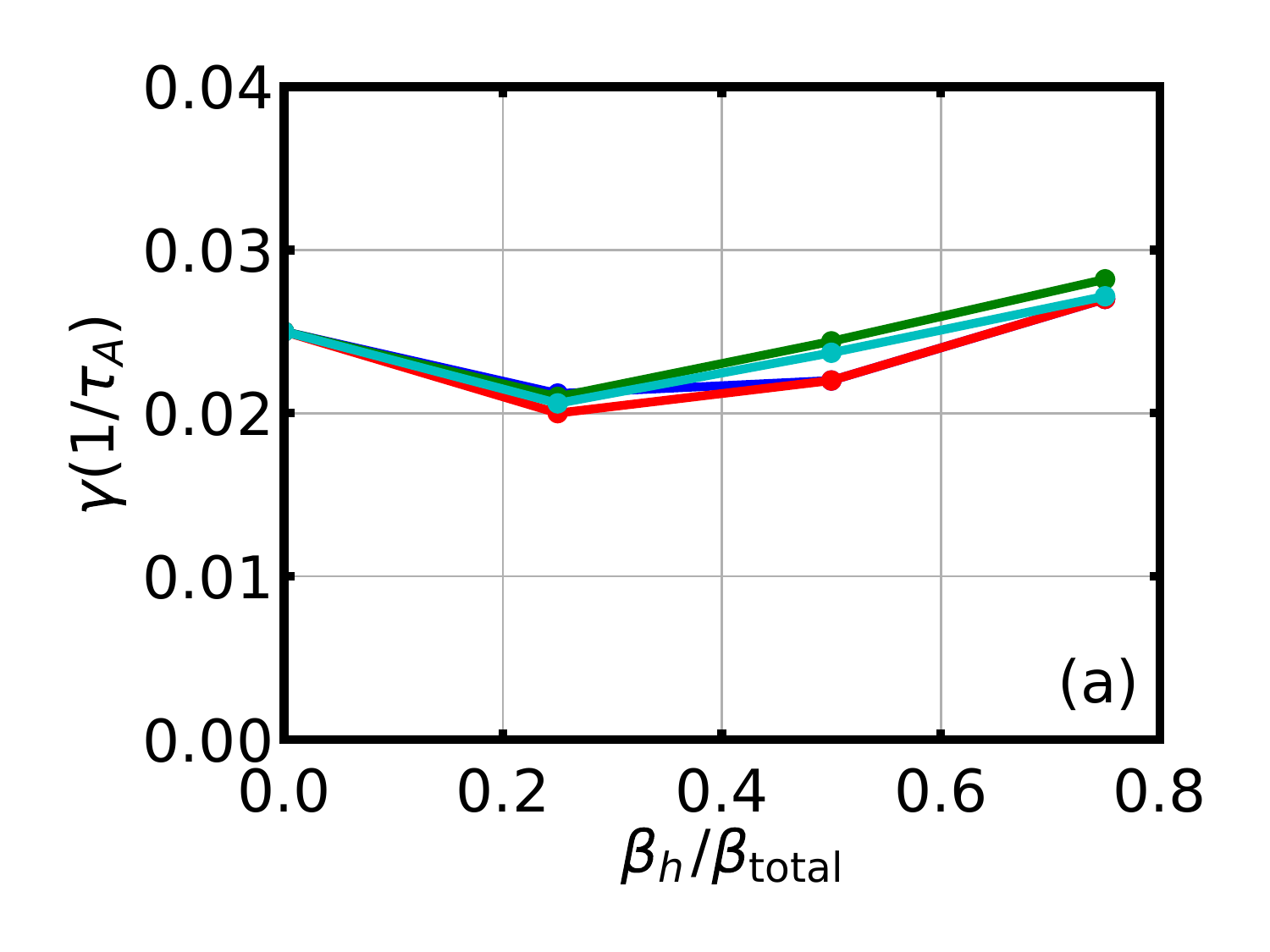}
		\includegraphics[width=0.39\linewidth]{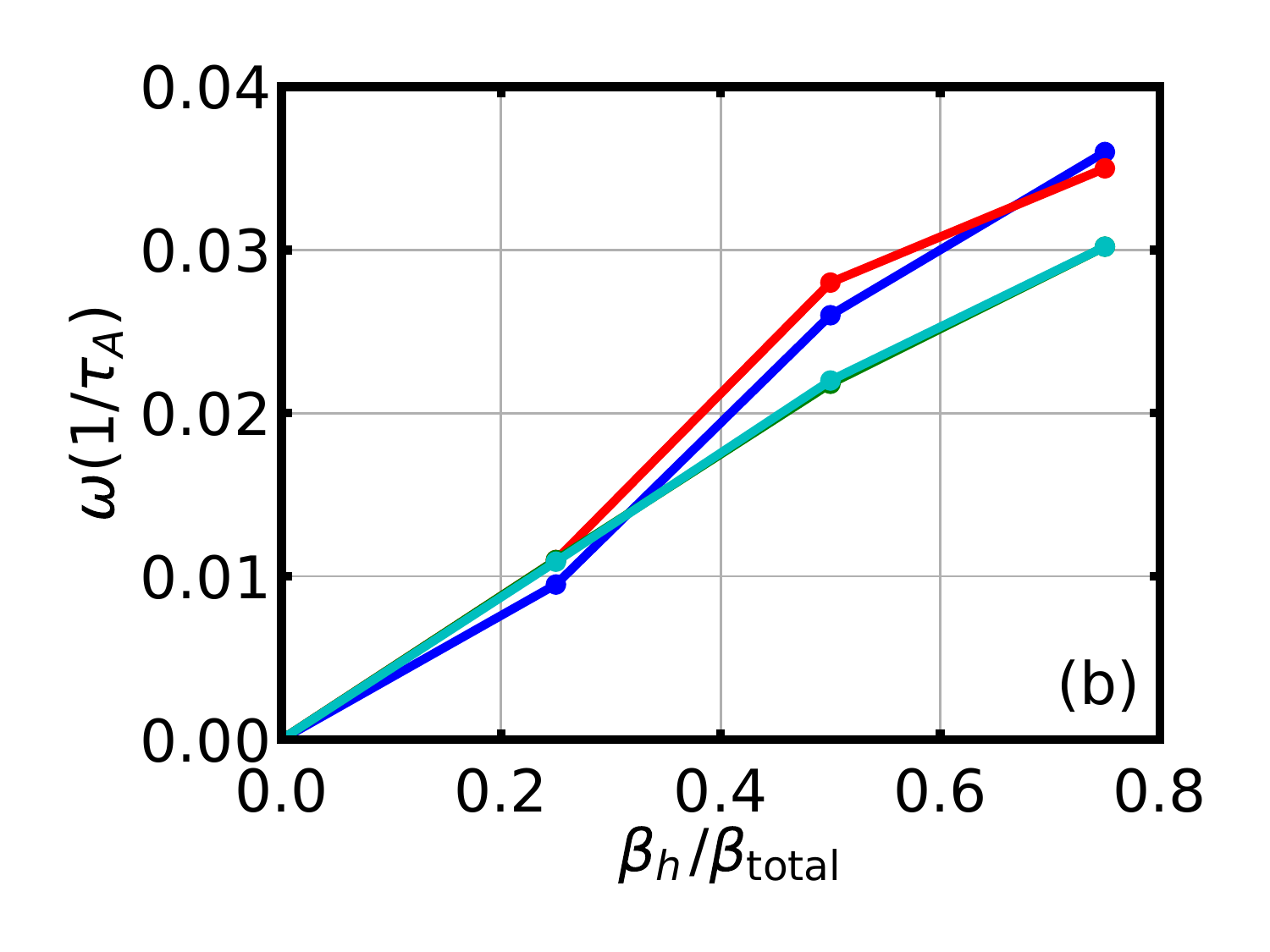}
		\raisebox{1.2\height}{\includegraphics[width=0.2\linewidth]{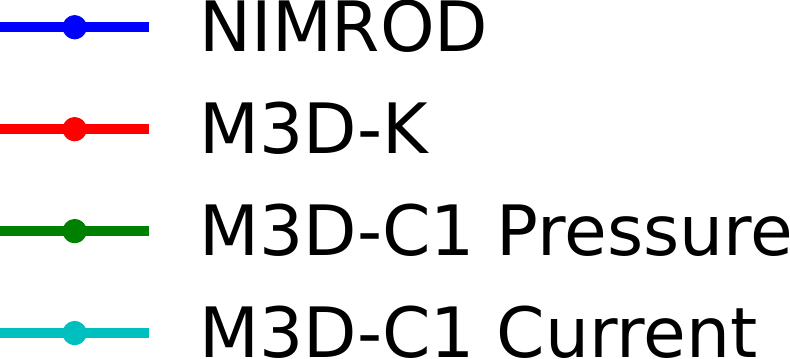}}
	\end{center}
	\caption{\label{fig:fishbone-benchmark}Simulation results of mode growth rate (a) and real frequency (b) as functions of \gls{ep} beta fraction of the $n=1$ fishbone. Blue line is the result of NIMROD\cite{kim_impact_2008}. Red line is the result of M3D-K\cite{fu_global_2006}. Green line is the result of M3D-C1-K using pressure coupling, and the cyan line is the result using current coupling.}
\end{figure}

The results of a linear simulation  with toroidal mode number $n=1$ are shown in \cref{fig:fishbone-benchmark}, including simulations using pressure coupling and current coupling schemes. The \gls{flr} effect was not included in the simulation. We can see that both the growth rate ($\gamma$) and the real frequency ($\omega$) agree well with the M3D-K and NIMROD results, except for the mode real frequency at large $\beta_h/\beta_{total}$. When $\beta_h/\beta_{total}$ increases from 0 to 0.75, the mode changes from an ideal \gls{mhd} kink mode to a fishbone mode with a finite real frequency due to the response of \glspl{ep}. The mode growth rate decreases as $\beta_h/\beta_{total}$ changes from 0 to 0.25, and then increases as $\beta_h/\beta_{total}$ changes from 0.25 to 0.75. The real frequency is zero with $\beta_h=0$ and increases almost linearly as $\beta_h$ increases. The results of simulations using pressure coupling and current coupling schemes are almost identical.  

\begin{figure}[h]
	\begin{center}
		\includegraphics[width=0.355\linewidth]{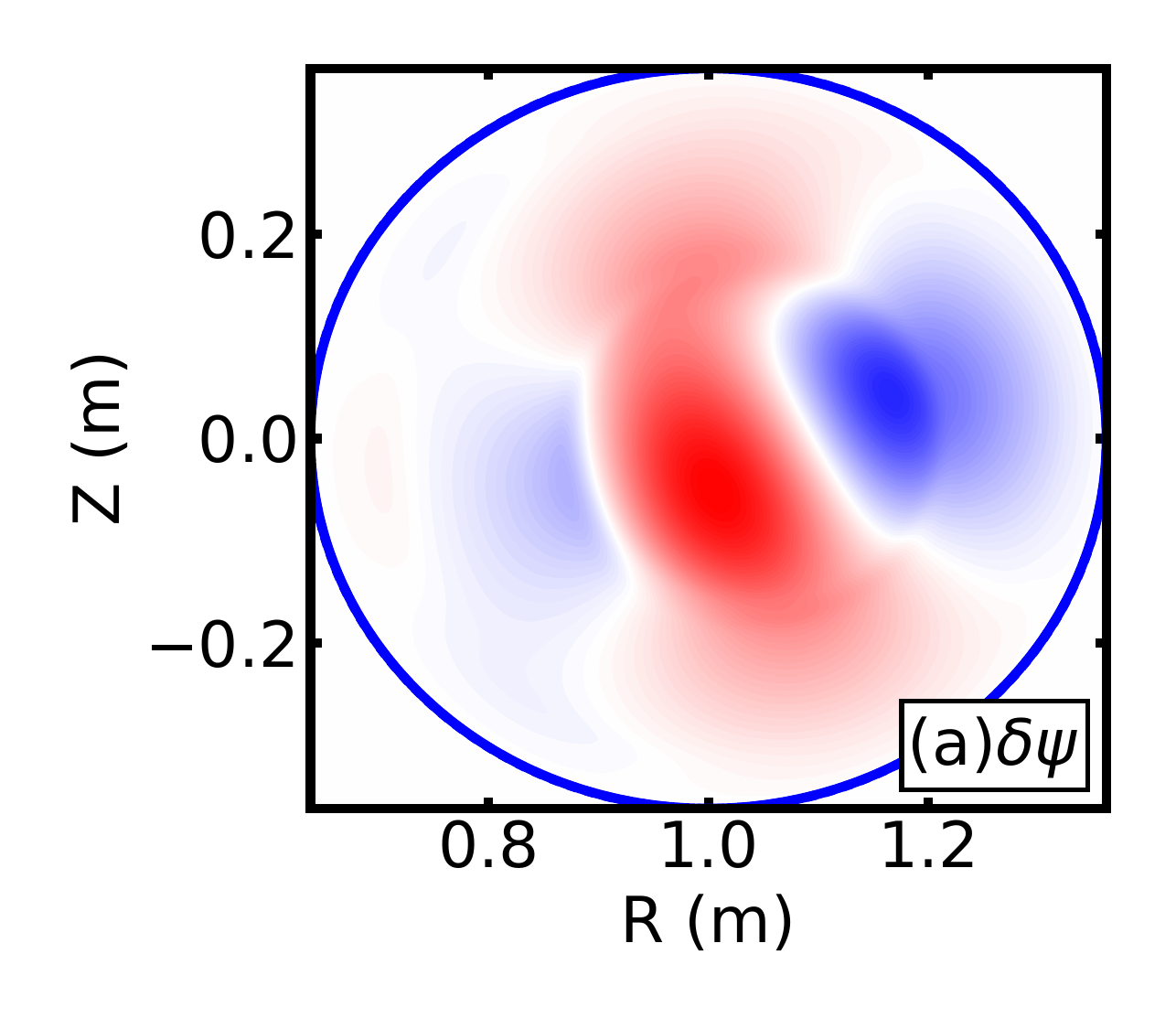}
		\includegraphics[width=0.28\linewidth]{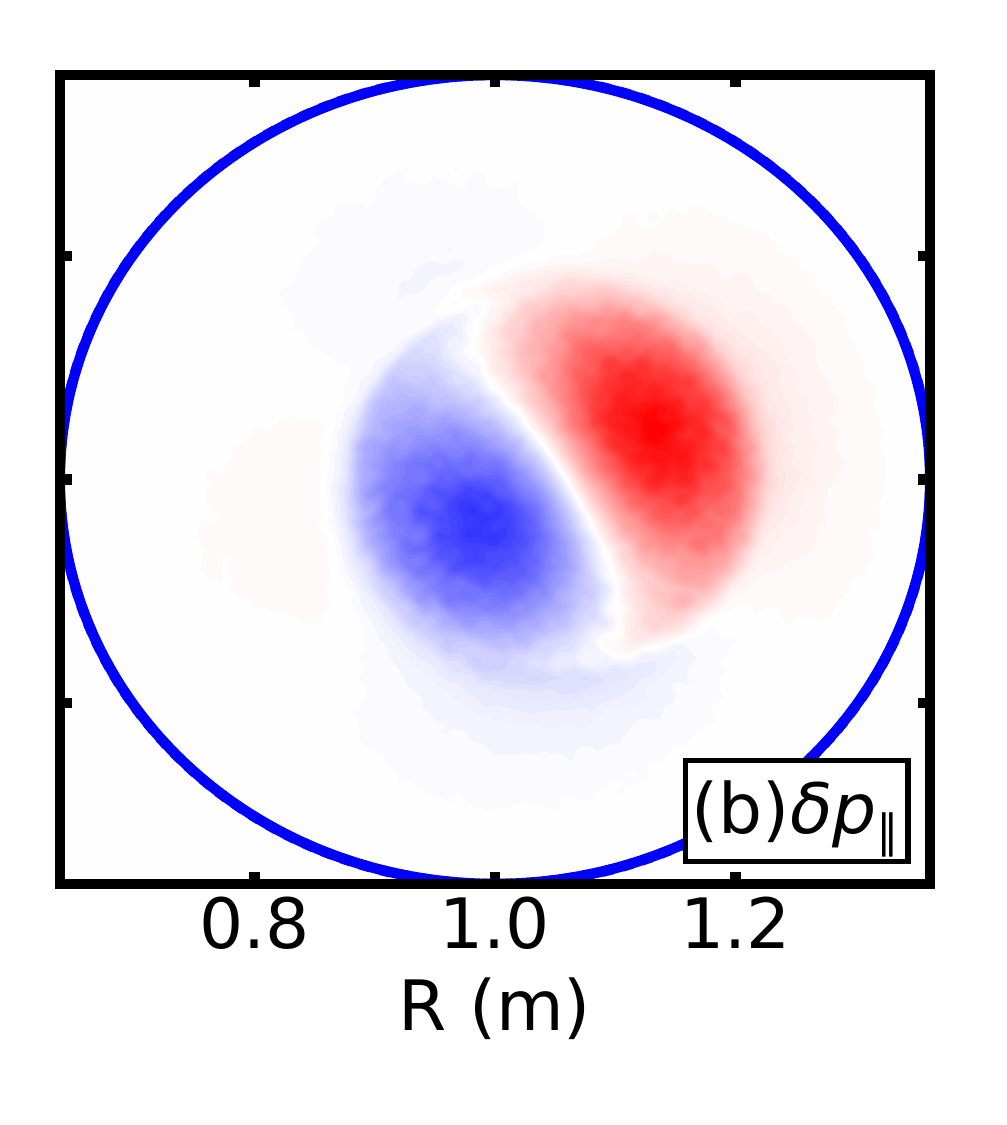}
		\includegraphics[width=0.28\linewidth]{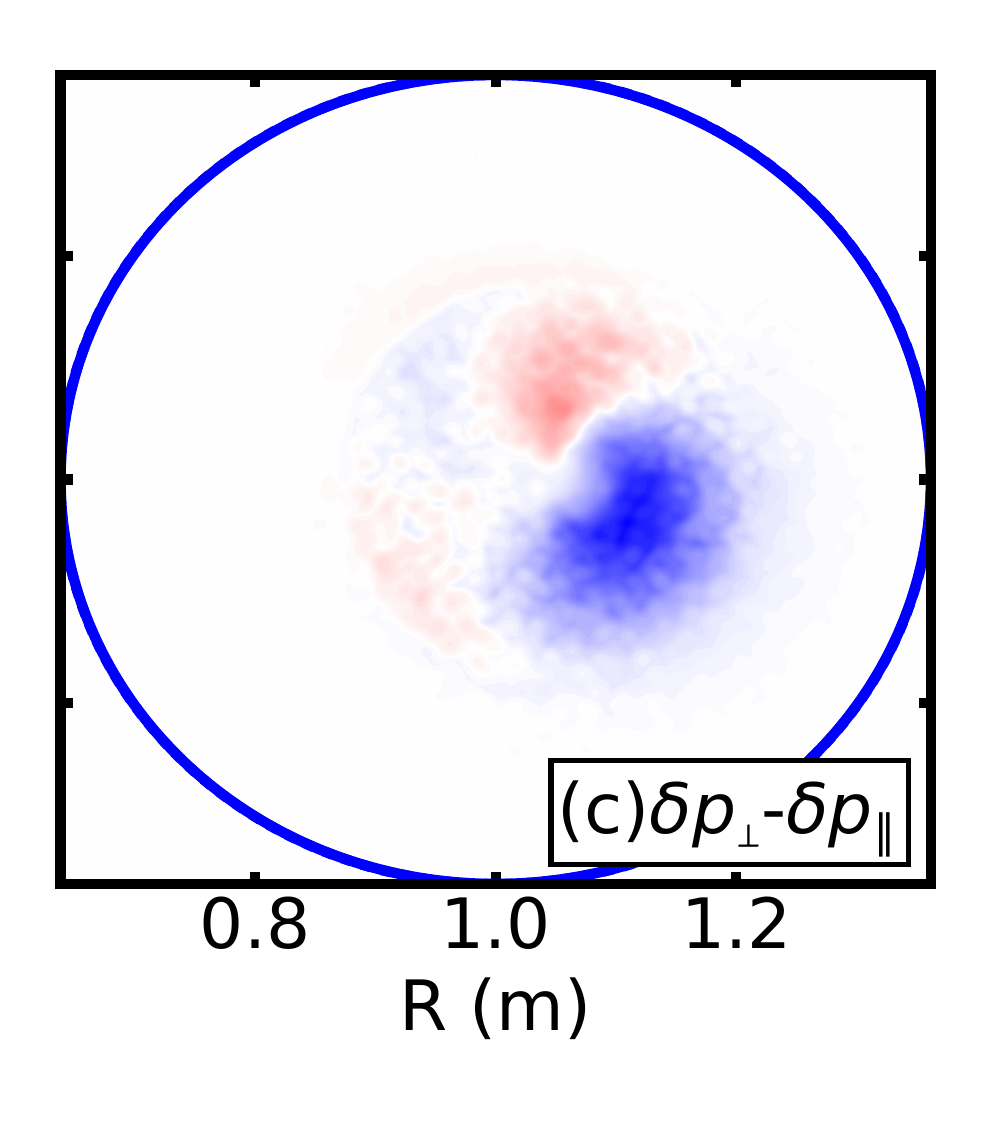}
		\raisebox{0.45\height}{\includegraphics[width=0.05\linewidth]{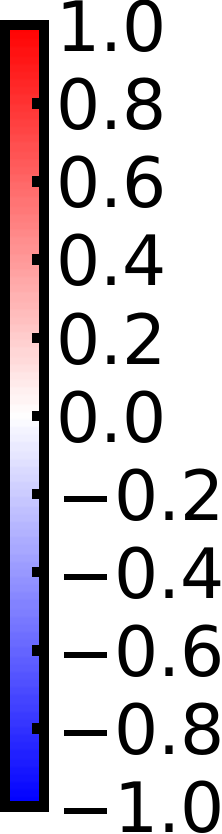}}
	\end{center}
	\caption{\label{fig:fishbone-psi}Structure of the perturbed poloidal flux $\delta \psi$ (a), the perturbed \gls{ep} parallel pressure $\delta p_\parallel$ (b) and the difference between the perturbed parallel and perpendicular \gls{ep} pressure $\delta p_\perp-\delta p_\parallel$ (c) from the $n=1$ linear fishbone simulation with $\beta_h/\beta_{total}=0.5$ using M3D-C1-K. The values are normalized according to the maximum absolute value.}
\end{figure}

The mode structure of the perturbed poloidal flux ($\delta\psi$), the perturbed \gls{ep} parallel pressure ($\delta p_\parallel$) and the difference between the perturbed parallel and perpendicular \gls{ep} pressure ($\delta p_\perp-\delta p_\parallel$) for a linear $n=1$ simulation with $\beta_h/\beta_{total}=0.5$ are shown in \cref{fig:fishbone-psi}. Note that the non-adiabetic response of \gls{ep} pressure ($\delta p_\perp-\delta p_\parallel$) is localized at the low-field-side, indicating that this pressure perturbation mostly comes from trapped particles through resonance with the fishbone mode. The particle pressure results have some noise because of the usage of the $\delta$ particle shape function and high-order polynomials as test functions. The mode structure results are consistent with the NIMROD simulation results in \cite{kim_impact_2008}.

\subsection{TAE simulation}

For \gls{tae} linear simulation we used the setup in \cite{konies_benchmark_2018}, which was also used for a NIMROD \gls{tae} simulation in \cite{hou_nimrod_2018}. The simulation was done in a large aspect ratio tokamak ($R=10$m, $a=1$m). The magnetic field on axis is $B_T=3$T. The bulk ions are hydrogen with a uniform density of $n_0=2\times 10^{19}$m$^{-3}$. The bulk plasma pressure is set to be constant to avoid pressure gradient driven modes, $p=6408$Pa. The safety factor profile is $q(r) = 1.71+0.16(r/a)^2$. Note that at $r=0.5a$ there is a rational surface $q=1.75$.

The energetic ions are deuterium and have a density profile given by
\begin{equation}
	n(s)=n_0 c_3\exp(-\frac{c_2}{c_1}\tanh\frac{\sqrt{s}-c_0}{c_2}),
\end{equation}
where $s=\psi_{t}/\psi_{t}(a)$ is the normalized toroidal flux. $n_0=1.4431\times 10^{17}$m$^{-3}$ is the \gls{ep} density at $s=0$. The coefficients $c_0=0.49123$, $c_1=0.298228$, $c_2=0.198739$ and $c_3=0.521298$. This \gls{ep} density profile has a large gradient at the rational surface $q=1.75$, which can drive \gls{tae}. The \glspl{ep} are initialized with a Maxwellian distribution in velocity space with a uniform temperature $T_f$.

\begin{figure}[h]
	\begin{center}
		\includegraphics[width=0.4\linewidth]{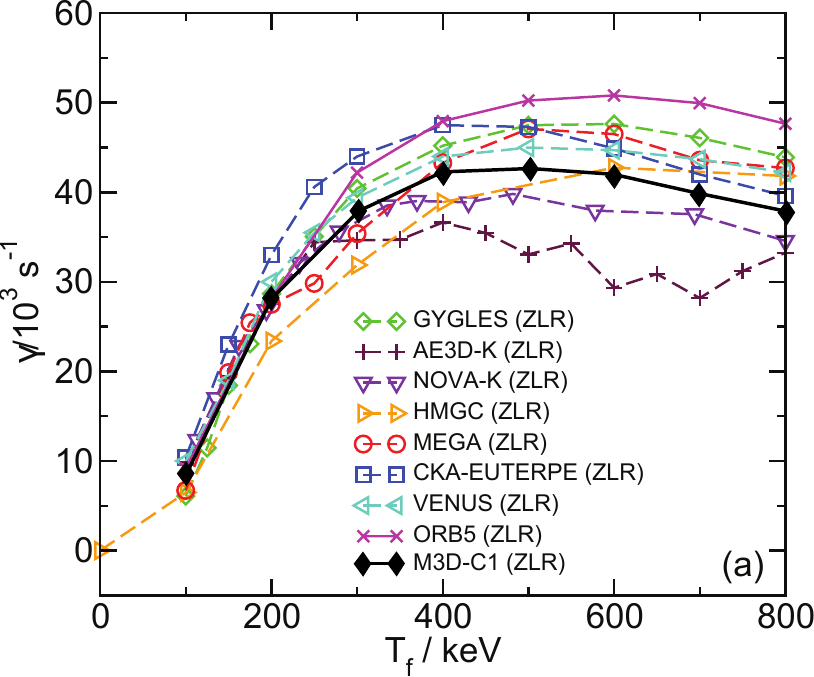}
		\includegraphics[width=0.4\linewidth]{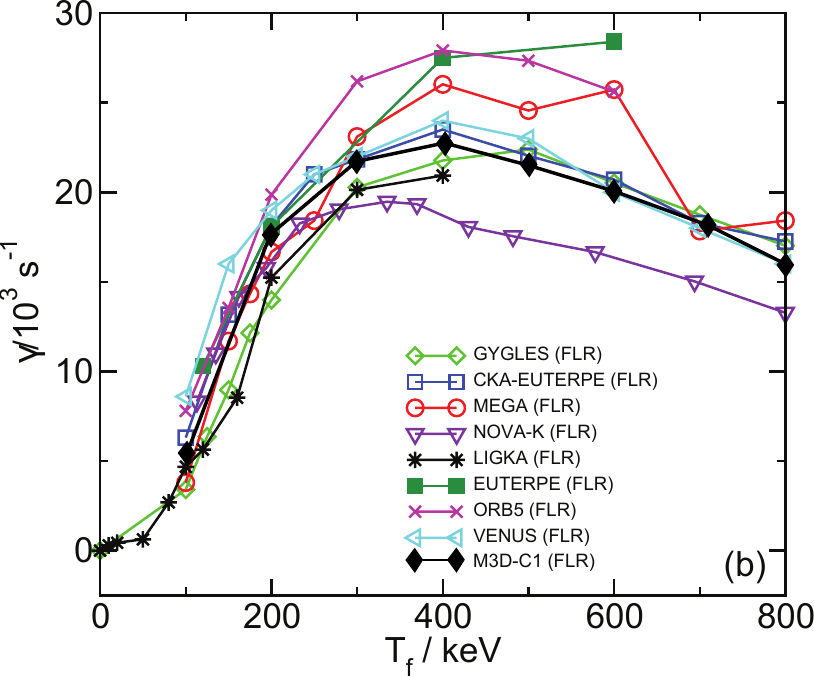}
		\includegraphics[width=0.35\linewidth]{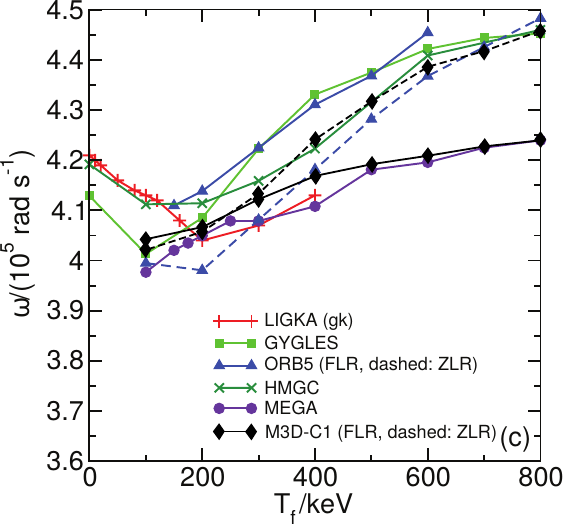}
	\end{center}
	\caption{\label{fig:dispersion-tae}Mode growth rates from calculations without \gls{flr} effects (a), with \gls{flr} effects (b) and the mode frequencies (c) as functions of $T_f$ for the linear $n=6$ \gls{tae} simulation. The black diamonds show the results from M3D-C1-K, on top of results from other codes presented in \cite{konies_benchmark_2018}.}
\end{figure}

The linear \gls{tae} simulation was done for $n=6$. The growth rates and frequencies of \gls{tae}s as functions of \gls{ep} temperature from the M3D-C1-K simulations are shown in \cref{fig:dispersion-tae}, including the results in the \gls{zlr} limit, and the results including \gls{flr} effect by taking 4-point gyro-averages. The results are plotted along with the simulation results from other codes which were benchmarked in \cite{konies_benchmark_2018}. We can see that the M3D-C1-K results are close to the results from gyrokinetic, hybrid-\gls{mhd} and eigenvalue codes. After including the \gls{flr} effect, the mode growth rates drop significantly for high $T_f$ cases as the \gls{ep} Larmor radius is large for those cases. The \gls{tae} frequencies also drop slightly with the \gls{flr} effect. We have done the simulations using both pressure and current coupling, and the results of mode growth rates and frequencies are equal.

\begin{figure}[h]
	\begin{center}
		\includegraphics[width=0.5\linewidth]{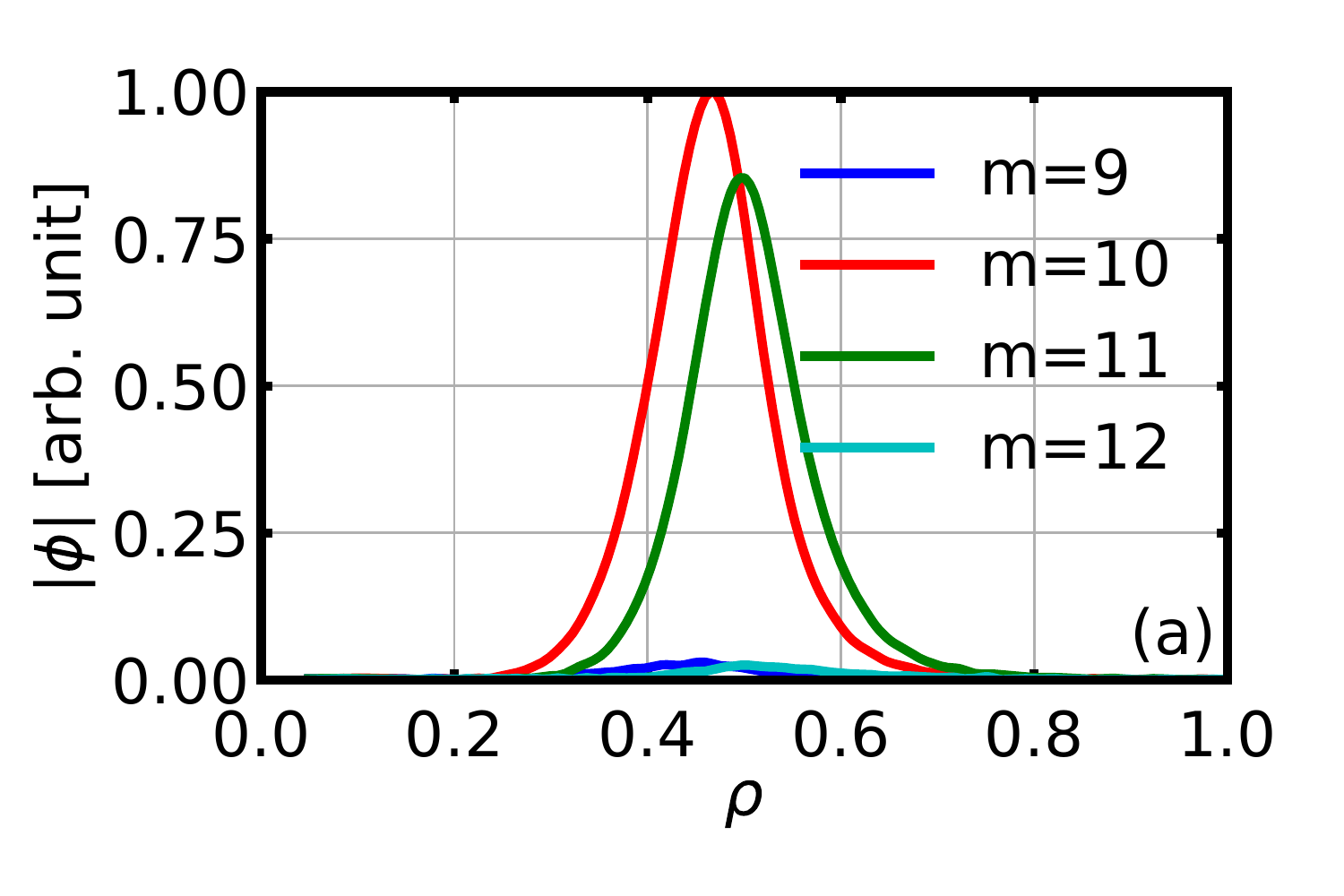}
		\raisebox{-0.03\height}{\includegraphics[width=0.4\linewidth]{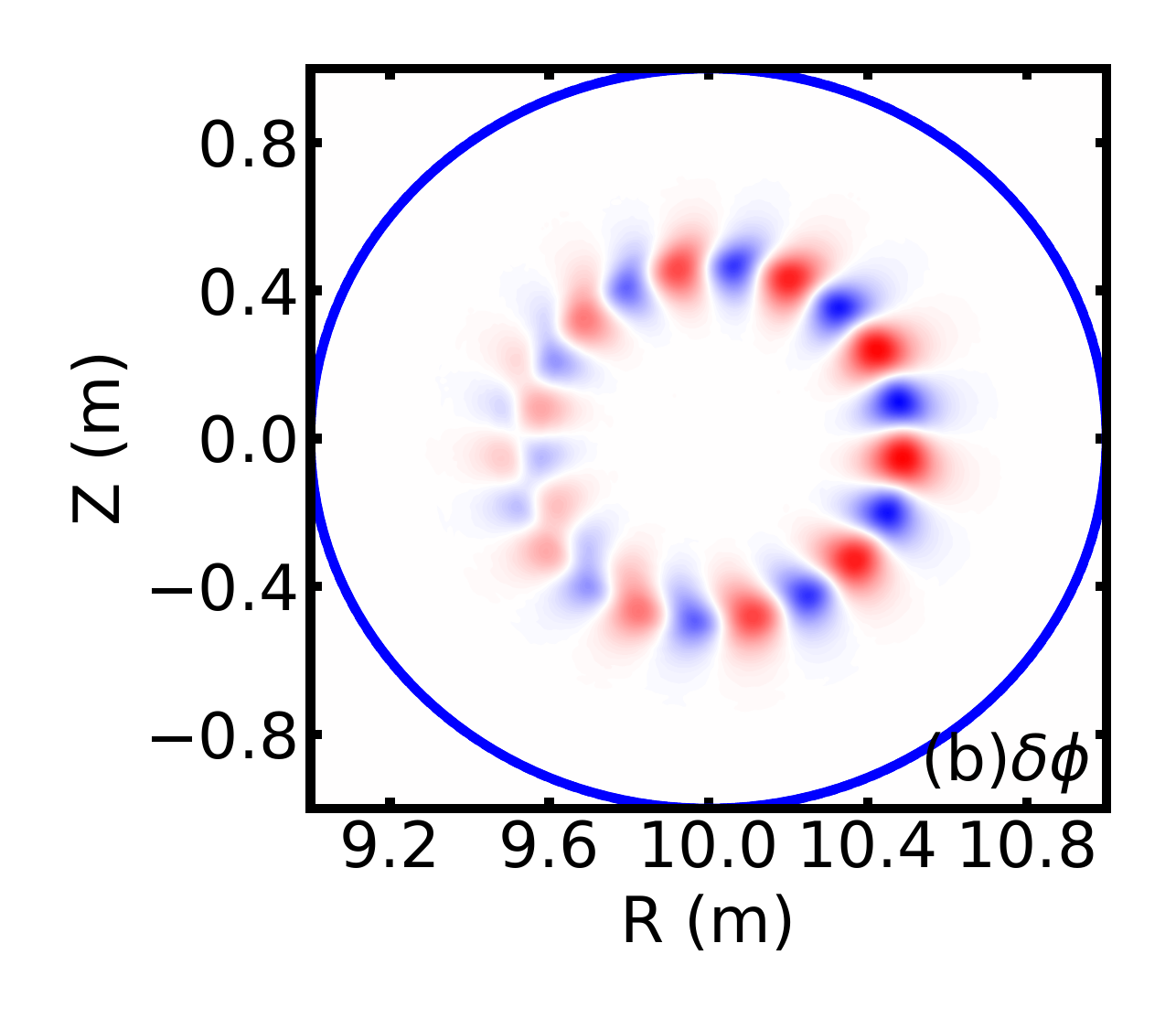}}
		\raisebox{0.22\height}{\includegraphics[width=0.07\linewidth]{fishbone-colorbar.pdf}}
	\end{center}
	\caption{\label{fig:tae-phi} (a) Poloidally averaged radial structure of perturbed poloidal vorticity $\delta \phi$ of different poloidal harmonics from the $n=6$ \gls{tae} simulation using M3D-C1-K. (b) Poloidal structure of $\delta \phi$. The values are normalized according to the maximum absolution value.}
\end{figure}

The mode structure of the perturbed poloidal vorticity ($\delta\phi$) from the M3D-C1-K simulation for $T_f=400$ keV including \gls{flr} effects is shown in \cref{fig:tae-phi}. The radial structure indicates that the mode is localized near the $r=0.5a$ rational surface and is dominated by $m=10$ and $m=11$ harmonics, which is consistent with the fact that the mode lies at the rational surface $q=1.75=0.5\times(10+11)/n$.

\subsection{RSAE simulation}

We also performed linear \gls{rsae} simulations in M3D-C1-K. For these simulations we used real tokamak geometry with plasma equilibrium and \gls{ep} distribution from experimental diagnostics. The equilibrium is obtained from DIII-D shot \#159243 at 805ms, during which the deuterium NBI is activated and a series of \glspl{rsae} were excited and measured \cite{collins_observation_2016,heidbrink_fast-ion_2017}. The simulation follows the setup in \cite{taimourzadeh_verification_2019}, in which a number of eigenvalue, gyrokinetic and hybrid-\gls{mhd} codes participated in a linear benchmark. The equilibrium fields, including the pressure profile, were read from the result of the equilibrium code kinetic EFIT, which takes into account the kinetic ion contribution in calculating the \gls{gs} equation. As shown in Fig.~3 in \cite{taimourzadeh_verification_2019}, the safety factor $q$ profile has a minimum point ($q_{min}=2.94$) at $\rho=0.4$ ($\rho$ is the normalized square root of toroidal flux). The \gls{ep} distribution is approximated by an isotropic Maxwellian distribution. Here we used the \gls{ep} density and temperature profile from kinetic EFIT, where the \gls{ep} pressure was estimated by subtracting the measured thermal pressure from the computed total pressure using the equilibrium reconstruction. The density and temperature profiles of both bulk plasma and fast ions used in the M3D-C1-K simulation have been carefully compared with the data used in \cite{taimourzadeh_verification_2019} to make sure they are in good agreement.

\begin{figure}[h]
	\begin{center}
		\includegraphics[width=0.5\linewidth]{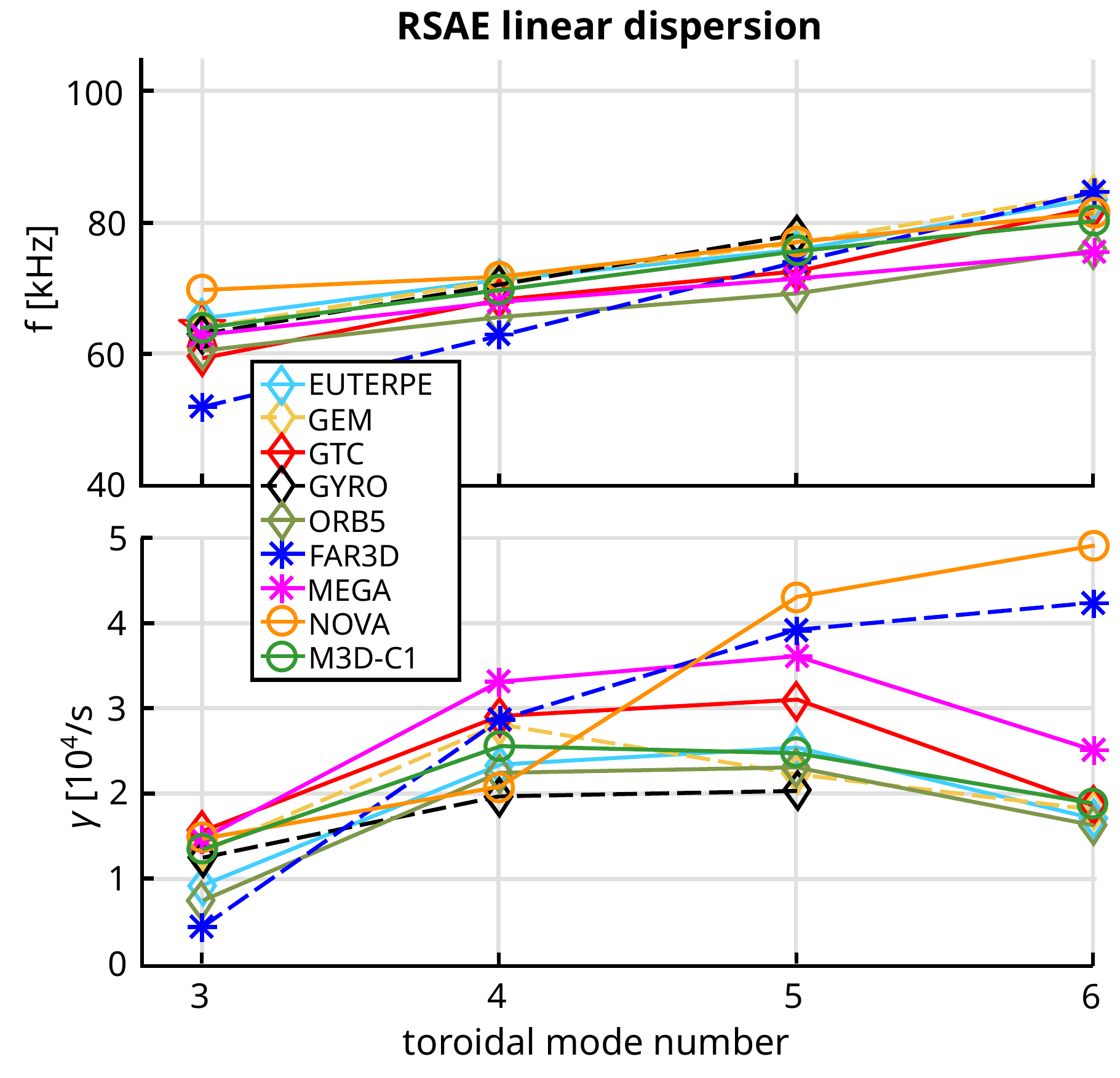}
	\end{center}
	\caption{\label{fig:dispersion-rsae}Mode frequencies ($f$) and growth rates ($\gamma$) from calculations with \gls{flr} effects for different $n$ numbers for the linear \gls{rsae} simulation. The green circles show the results from M3D-C1-K, on top of results from other codes presented in \cite{taimourzadeh_verification_2019}}
\end{figure}

\begin{figure}[h]
	\begin{center}
		\includegraphics[width=0.5\linewidth]{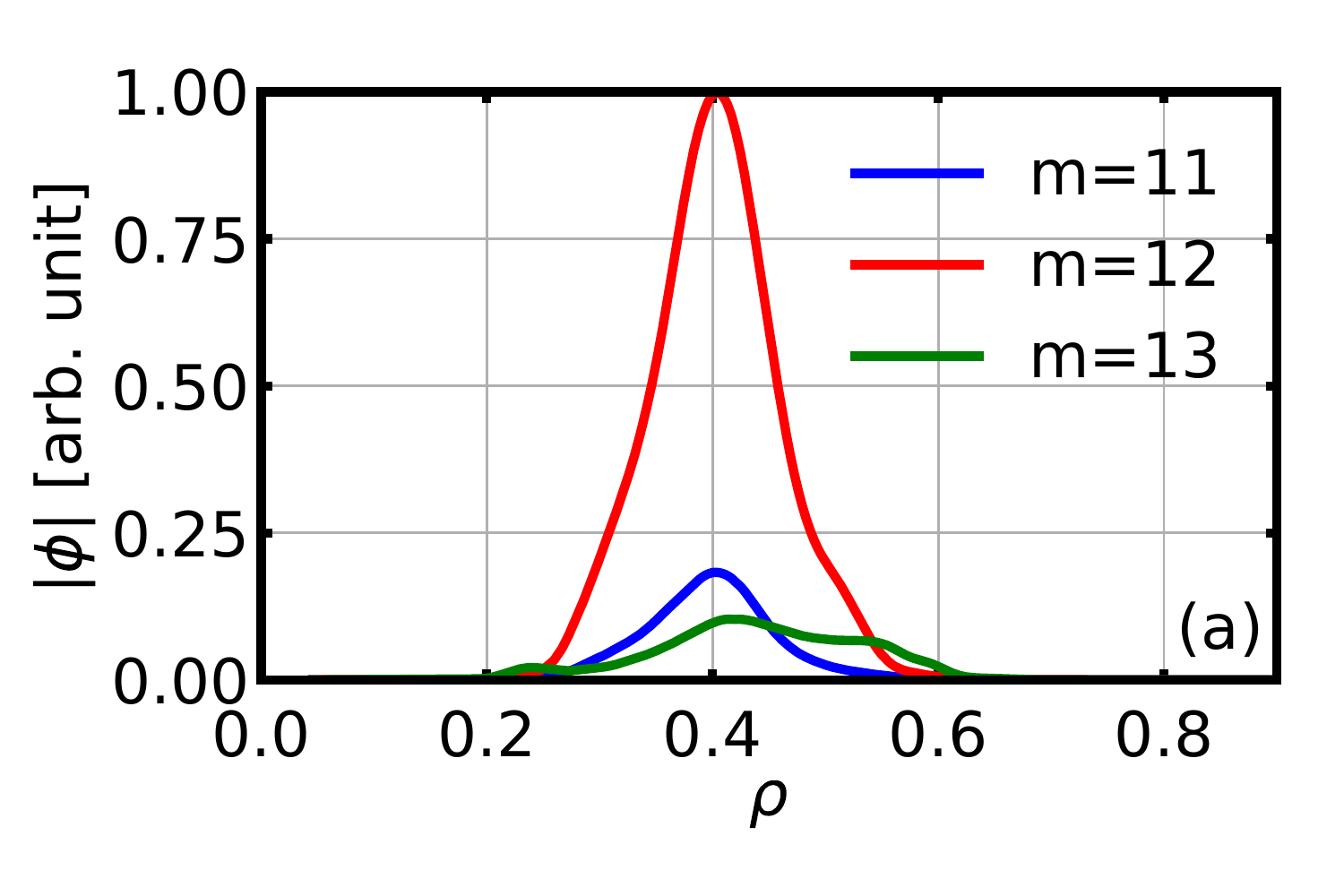}
		\raisebox{-0.1\height}{\includegraphics[width=0.4\linewidth]{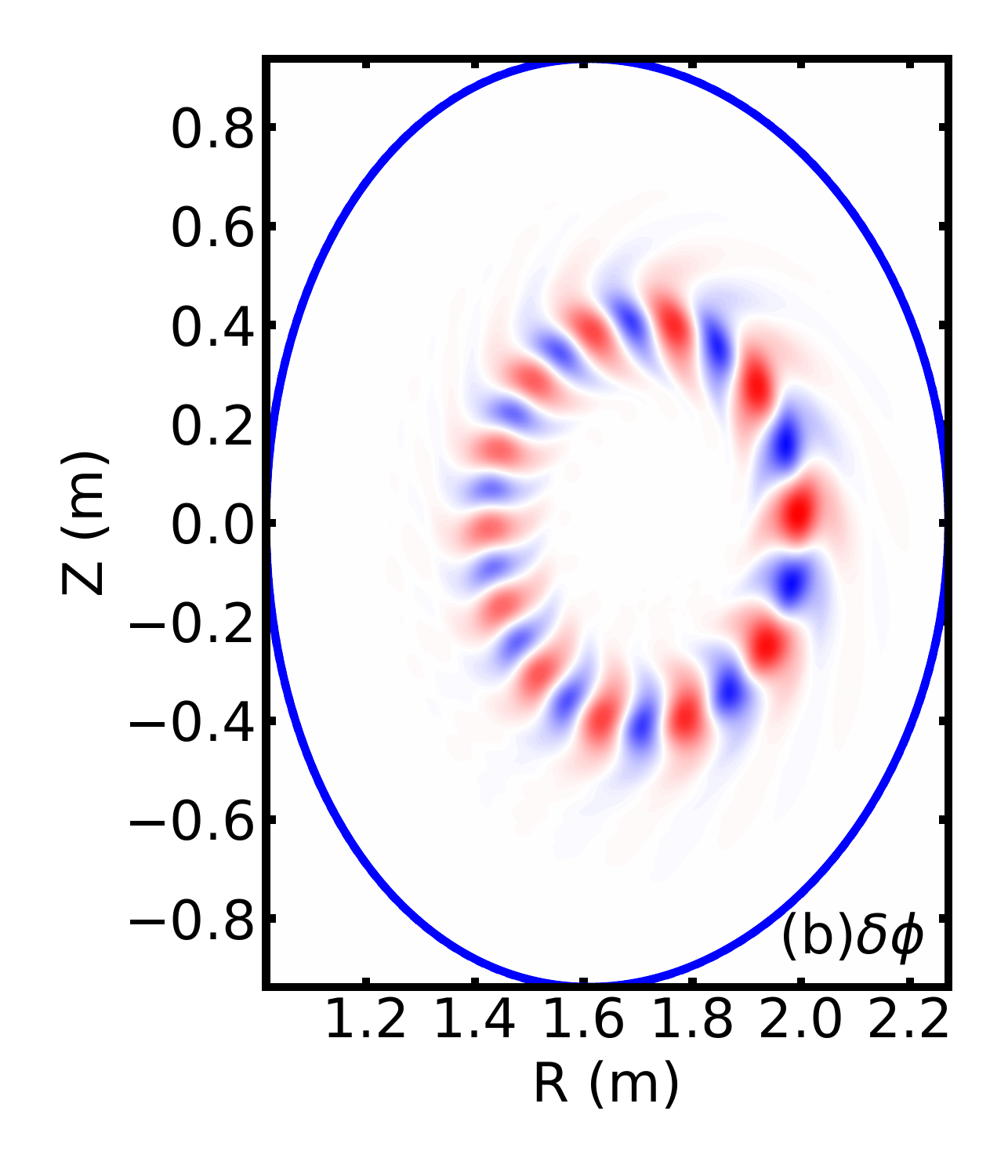}}
		\raisebox{0.2\height}{\includegraphics[width=0.08\linewidth]{fishbone-colorbar.pdf}}
	\end{center}
	\caption{\label{fig:rsae-phi}(a) Poloidally averaged radial structure of perturbed poloidal vorticity $\delta \phi$ of different poloidal harmonics of the $n=4$ \gls{rsae} simulation using M3D-C1-K. (b) Poloidal structure of $\delta \phi$. The values are normalized according to the maximum absolution value.}
\end{figure}

Using this equilibrium, we did linear simulations using M3D-C1-K for $n=3-6$. The results of the \gls{rsae} real frequencies and growth rates are plotted in \cref{fig:dispersion-rsae} along with the results from other codes presented in \cite{taimourzadeh_verification_2019}. The M3D-C1-K results agree well with results from the other initial value \gls{mhd} and gyrokinetic codes. The mode frequencies increase as the $n$ number, while the growth rate is largest for $n=4$ and 5. For these simulations we include the \gls{flr} effect, and we found that \gls{flr} can lead to a decrease of the mode growth rate similar to what we found for the \gls{tae} simulation. The mode structure of the $n=4$ \gls{rsae} simulation is shown in \cref{fig:rsae-phi}, including the radial structure of different $m$ harmonics of $\delta \phi$ and the 2D poloidal structure. The perturbed field is localized near the $q=q_{min}$ flux surface and is dominated by the $m=12$ component, which is consistent with the \gls{rsae} physics ($q_{min}\approx m/n$) and in agreement with the results in \cite{taimourzadeh_verification_2019} from the other codes.

\section{Conclusions}
\label{sec:conclusion}

In this paper we have introduced the new code M3D-C1-K, which was developed based on the M3D-C1 \gls{mhd} code with particle simulation for the kinetic effects. The particles are described using markers, which are pushed using a new slow manifold Boris algorithm. This new algorithm can provide good conservation properties for long time simulations. In addition, it can simplify the field evaluation calculation and speed up the particle pushing. The particle simulation is interfaced with the \gls{mhd} code by calculating the moments of the particles using the $\delta f$ method, and then coupled into the \gls{mhd} equations through pressure or current. The particle pushing code has been ported to run on \glspl{gpu}, which gives a 11 times speed up compared to the \gls{cpu} version. Both the linear fishbone simulations and the linear Alfvén mode simulations, including \gls{tae} and \gls{rsae}, have been conducted, and the results agree well with previous results from other codes.

M3D-C1-K is based on M3D-C1, which utilizes the semi-implicit method to do \gls{mhd} calculations with large timesteps. To fit the kinetic part into this framework, we integrate the \gls{mhd} and particle equations separately, and introduce subcycles for particle pushing. Given that the \gls{mhd} equations are still evolved using a large timestep which is not limited by the \gls{cfl} condition, and particle pushing on \glspl{gpu} is very fast, we believe that M3D-C1-K is suitable for simulation of long-time \gls{mhd} phenomena involving kinetic effects, including the nonlinear evolution of \gls{ep}-driven Alfvén modes with frequency chirping and mode coupling, and kink or tearing modes interacting with \glspl{ep}. For these kind of simulations, the computation time spent on the \gls{mhd} calculation on \glspl{cpu} and particle pushing on \glspl{gpu} are comparable. For phenomena involving wave-particle interaction over short timescales, such as \glspl{gae} or \glspl{cae}, small \gls{mhd} timesteps are required which can make the \gls{mhd} calculation take most of the computation time. In order to better simulate these kinds of problems, we plan to further optimize the \gls{mhd} calculation and have it utilize \glspl{gpu}.

The new slow manifold Boris algorithm used in the code was originally developed to preserve physical structures and conserve constants of motion, which can improve the credibility of long time simulations. As discussed in \cref{sec:slow-manifold}, this advantage is not significant for a typical \gls{ep} simulation of only hundreds of milliseconds as \gls{rk4} can provide similar order of absolute numerical error. For longer time simulations the benefit of the Boris algorithm can be more significant. In addition, this advantage can be more important for simulating particles with large parallel velocities such as high energy electrons. These electrons can be generated through inductive electric fields as runaway electrons, or through external current drive with plasma waves, and can interact with \gls{mhd} modes. Given that the high-energy electrons can have velocities close to the speed of light, it is important to have a particle pushing algorithm that can conserve the toroidal momentum and keep the shape of the particle's orbit, as discussed in \cite{guan_phase-space_2010,liu_self-consistent_2021}. The slow manifold Boris algorithm therefore is a good candidate for doing nonlinear \gls{mhd} simulation with energetic electrons, and will be discussed in future studies.

\section*{Acknowledgments}

We would like to thank Yasushi Todo, Andreas Bierwage, Elena Belova, Nikolai Gorelenkov, Xin Wang, Roscoe White, Jin Chen, Zhihong Lin and Amitava Bhattacharjee for fruitful discussion. We would like to thank Seegyoung Seol and Mark Shephard of the Scientific Computation Research Center (SCOREC) group at Rensselaer Polytechnic Institute (RPI) for the implementation and support of
unstructured meshing capabilities in M3D-C1. This work was supported by US Department of Energy grants DE-AC02-09CH11466. This research used the Traverse cluster at Princeton University and AiMOS cluster of Center for Computational Innovations (CCI) at RPI. It also used the Summit cluster of the Oak Ridge Leadership Computing Facility at the Oak Ridge National Laboratory, which is supported by the Office of Science of the U.S. Department of Energy under Contract No. DE-AC05-00OR22725.

\bibliography{paper}

\end{document}